\documentclass[twocolumn,superscriptaddress,longbibliography,
aps,pra,noshowpacs]{revtex4-2}

\usepackage{amsmath}
\usepackage{amsfonts}
\usepackage{amssymb}
\usepackage{ulem}
\usepackage{xcolor}

\usepackage{graphicx}
\usepackage{hyperref}
\hypersetup{
	colorlinks=true,
	citecolor=blue,
	linkcolor=blue,
	urlcolor=blue}

\begin{document}

\title[]{Temperature driven false vacuum decay in coherently coupled Bose superfluids}

\author{Paniyanchatha Moolayil Sivasankar}
\email{ud23002@students.iitmandi.ac.in }
\affiliation{School of Physical Sciences, Indian Institute of Technology Mandi, Mandi-175075 (H.P.), India}

\author{Franco Dalfovo}
\email{franco.dalfovo@unitn.it }
\affiliation{Pitaevskii BEC Center, CNR-INO and Dipartimento di Fisica, Universit\`a di Trento, I-38123 Trento, Italy}

\author{Alessio Recati}
\email{alessio.recati@unitn.it}
\affiliation{Pitaevskii BEC Center, CNR-INO and Dipartimento di Fisica, Universit\`a di Trento, I-38123 Trento, Italy} 
\affiliation{
Trento Institute for Fundamental Physics and Applications, INFN, I-38123 Trento, Italy}

\author{Arko Roy}
\email{arko@iitmandi.ac.in}
\affiliation{School of Physical Sciences, Indian Institute of Technology Mandi, Mandi-175075 (H.P.), India}

\begin{abstract}
The relaxation of a quantum field from a metastable state (false vacuum) to a stable one (true vacuum), also known as false vacuum decay, is a fundamental problem in quantum field theory and cosmology. We study this phenomenon using a two-dimensional interacting and coherently coupled Bose-Bose mixture, a platform that has already been employed experimentally to investigate false vacuum decay in one dimension. In such a mixture, it is possible to define an effective magnetization that acts as a quantum field variable. Using the Stochastic Projected Gross-Pitaevskii equation (SPGPE), we prepare thermal equilibrium states in the false vacuum and extract decay rates from the magnetization dynamics. The decay rates show an exponential dependence on temperature, in line with the thermal theory of instantons. Since the SPGPE is based on complex scalar fields, it also allows us to explore the behavior of the phase, which turns out to become dynamic during decay. Our results confirm the SPGPE as an effective tool for studying coupled magnetization and phase dynamics and the associated instanton physics in ultracold quantum gases.
\end{abstract}

\maketitle

\section{\label{sec:intro}Introduction}
False vacuum decay (FVD) describes the relaxation of a quantum system from a metastable false vacuum (FV) to a lower-energy true vacuum (TV). In his seminal work, Coleman showed that this transition proceeds via the stochastic nucleation of true vacuum bubbles within the false vacuum background, triggered by quantum fluctuations~\cite{Coleman_77,Callan_77}. Such a decay finds many applications in different areas of science, including early dynamics of our universe~\cite{Kobzarev_74,Coleman_77,Callan_77,Coleman_80,M.hindmarsh_21,Burda_15}, high-energy physics~\cite{MSTONE_77,SHAPOSHNIKOV_87,Batini_24,Hirvonen_25}, protein folding~\cite{Baldwin_11,Ghosh_20} and condensed matter systems~\cite{DWOxtoby_92,Debenedetti_01, R.Coldea_10}. However, the strongly non-perturbative character of this process has long hindered its direct exploration.

Recent advances in quantum simulation have opened new routes for emulating false vacuum decay in controllable platforms. Proposals for analog realizations range from spin systems~\cite{Rutkevich_99,A.sinha_21,Lagnese_24,johansen_25,luka_pavesic_25,daan_maertens_25,Borla_26} and digital simulators~\cite{Abel_21,Vodeb_25} to ultra-cold atomic gases~\cite{Fialko_17,braden_18,Billam_20,Billam_21,Billam_22,zenesini_24,Darbha_24,K.brown_25,jenkins_25}, the latter being particularly well suited owing to their isolation from the environment, tunable interactions, and real-time access to non--equilibrium dynamics. 

\begin{figure}
    \centering
    \includegraphics[width=0.8\linewidth]{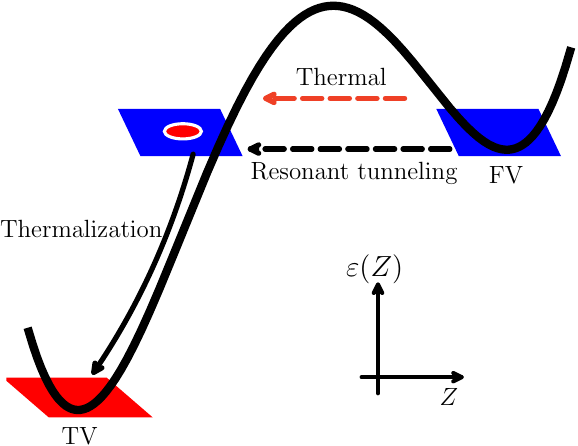}
    \caption{Schematic illustration of false vacuum decay through the energy landscape. Here  $\varepsilon$ is the energy and $Z$ is the magnetization. The metastable false vacuum corresponds to the fully polarized ${Z}=+1$ state on the right (blue plane). It decays by nucleating bubbles (i.e., regions of opposite polarization) which then evolve towards the true vacuum $Z=-1$ state on the left (red plane) through thermalization processes. 
    }
    \label{fig:fvd_sketch}
\end{figure}

In particular, coherently coupled Bose–Einstein condensates \cite{Alessio_22} have recently allowed for the experimental observation of false vacuum decay~\cite{zenesini_24,cominotti_25}.
Despite the apparent simplicity of the system, it hosts many phenomena ranging from internal Josephson dynamics~\cite{T.Zibold_10,farolfi_2021_torque,Farolfi_21}, quark confinement~\cite{MinoruEto_18} to magnetism~\cite{Alessio_22,Cominotti_23}. The system is a gas of atoms dressed by a Rabi coupling between two hyperfine levels that, at low enough temperature, forms a Bose-Einstein condensate with a spinor order parameter. Due to the lack of $SU(2)$ symmetry in the atomic  interaction strengths, the system behaves like an anisotropic ferromagnetic material, and it exhibits a para- to ferromagnetic (easy-axis) transition characterized by the spontaneous $\mathbb{Z}_2$ ordering of the longitudinal spin component. This transition can be described by the transverse field Ising model and is governed by the ratio between the Rabi coupling and the anisotropy in the intra- and inter-species interaction strengths~\cite{Cominotti_23}.  

In the ferromagnetic phase, the energy landscape as a function of the magnetization (i.e., the relative imbalance of the two components)  forms a symmetric double-well potential. This symmetry can be deliberately broken by introducing a finite detuning  between the hyperfine levels resonant condition and the oscillating radio-frequency field that couples the two components. This detuning renders the energy landscape an asymmetric double well, as schematically shown in Fig.~\ref{fig:fvd_sketch}, and it is an experimentally tunable parameter that can be used to prepare the system under the appropriate conditions to observe false vacuum decay

Theoretical predictions for the decay rate have been obtained in the past by using the instanton theory, corresponding to the imaginary time solutions of the  equations of motion of the field~\cite{Coleman_77,Callan_77}. A.D.Linde provided a finite temperature extension of the instanton theory showing that it can greatly enhance the decay rate~\cite{Linde_83}. Although such a semiclassical picture captures the basic mechanism of FVD, many of the details remain to be verified. Furthermore, the effects of temperature have only been observed experimentally very recently, in a quasi-one-dimensional trapped system~\cite{cominotti_25}. 

In this work, we use the Stochastic Projected Gross–Pitaevskii formalism to theoretically investigate FVD in a two-dimensional homogeneous system. This theory treats Bose-Einstein condensates at the classical-field level, including temperature effects through fluctuations of a $c$-field governed by a Gross-Pitaevskii-like equation and coupled to a thermal bath of incoherent atoms~\cite{Stoof_01,Proukakis_08,Blakie_08}. In this way, we simulate the FVD process in configurations that are within the reach of state-of-the-art experiments. We qualitatively compare our results with the prediction of finite temperature instanton theory.    

The article is organized as follows. In Sec.~\ref{sec:Model system} we introduce our model system and discuss how it can be used within a stochastic framework to emulate FVD. In Sec.~\ref{sec:FVD}, we discuss the theory of FVD and present our numerical results.


\section{\label{sec:Model system}Theoretical framework and initial state preparation}

\subsection{Zero temperature energy landscape}

We consider a dilute, weakly-interacting homogeneous two-component atomic Bose gas in two-dimensions, confined in a square box of area $\mathcal{A}=L_x \times L_y$, with periodic boundary conditions. The two components correspond to atoms in different hyperfine states $|1\rangle$ and $|2\rangle$.  An oscillating radio frequency (rf) field of strength $\Omega$, which is detuned from resonance by $\delta$, creates a linear superposition of the two states. The two states of interest are typically separated by energies in the MHz–GHz range, as they belong to the same or different hyperfine manifolds. This makes them directly accessible to single-photon rf radiation, enabling controlled and coherent coupling between the states in a straightforward and experimentally well-established manner~\cite{baroni_24,farolfi_phd21}. By mapping the system to a spin-$1/2$ model, one can define the magnetization $Z=(N_1-N_2)/N$,  where $N_i$ is the number of atoms in component $i=1,2$, and  $N=N_1+N_2$ is the total number. 
Atoms interact via a two-body contact potential, with intra- and inter-species constants $g_{11}$, $g_{22}$, $g_{12}$. We assume a symmetric mixture with $g_{11}=g_{22}=g$, and define $G=(g+g_{12})/2$ as the density interaction strength, and $\kappa=(g-g_{12})/2$ as the spin interaction strength.

Before addressing the behavior at finite temperature, where the gas is subjected to local fluctuations in density and phase, it is useful to recall what is expected at $T=0$. In this case, the densities $n_1$ and $n_2$ of the two components, as well as the total density $n=n_1+n_2$, are constant and the mean-field (grand-canonical) energy density is given by~\cite{Alessio_22,Arko_23}

\begin{eqnarray}
\varepsilon(n,Z,\varphi) &=& \tfrac{1}{2}Gn^2+\tfrac{1}{2}\kappa n^2Z^2- n Z \delta
\nonumber\\
&-& \Omega n \sqrt{1-Z^2}\cos\varphi -\mu n,
\label{eq:energy_density}
\end{eqnarray}
where $\varphi$ is the relative phase of the two components and $\mu$ is the chemical potential.
The ground state corresponds to $\cos \varphi=1$ and is obtained by minimizing $\varepsilon(n,Z)$ with respect to $Z$ and $n$, which yields
\begin{eqnarray}
Z \kappa n-\delta+\frac{\Omega Z}{\sqrt{1-Z^2}}&=&0,\nonumber\\
Gn-\frac{\Omega}{\sqrt{1-Z^2}}&=&\mu.
\label{chemical_potential}
\end{eqnarray}
These equations can be solved numerically by first determining $Z$ and then evaluating $\mu$.

For $\delta = 0$, Eq.~(\ref{eq:energy_density}) describes a degenerate mean-field energy landscape that supports two distinct phases. When $\Omega < |\kappa| n$, the system exhibits a polarized ferromagnetic phase, whereas for $\Omega > |\kappa| n$ it remains in an unpolarized paramagnetic phase. A finite detuning $\delta$ breaks the underlying $\mathbb{Z}_2$ symmetry, resulting a non-degenerate energy landscape. The detuning term energetically biases the minima at $Z \approx \pm 1$ by an amount $2n\delta$, leading to the coexistence of local and global minima.

\subsection{\label{sec:SGPE}Stochastic Projected Gross-Pitaevskii equation}

Given the energy landscape, our goal is to prepare the system at equilibrium in the global free energy minimum at a finite temperature $T$, then modify the detuning so that the system is transferred into a false vacuum state with thermal fluctuations. For this we use the Stochastic Projected Gross-Pitaevskii equation (SPGPE)
\begin{eqnarray}
i \hbar \frac{\partial \psi_i}{\partial t}  
&=& \hat{{\mathcal{P}}} \Bigg\{ (1 - i\gamma) \Bigg[ \Bigg( - \frac{\hbar^2 \nabla^2}{2 m} + g |\psi_i|^2 + g_{12} |\psi_{3 - i}|^2 \nonumber\\
&&  + (-1)^i \delta - \mu \Bigg) \psi_i - \Omega \psi_{3-i} \Bigg] + \eta_i \Bigg\} , 
\label{eq:SGPE-eqn}
\end{eqnarray}
which corresponds to the extension to multi-component systems~\cite{A.bradley_14,Liu_16,miki_ota_18,Arko_21,Arko_23} of the standard single-component SPGPE~\cite{Stoof_01,Proukakis_08,Blakie_08}. 
Here, $\psi_i(\mathbf{x},t)=\sqrt{n_i(\mathbf{x},t)} e^{i\phi_i(\mathbf{x},t)}$ are coherent $c$-fields in the coordinate space $\mathbf{x}=(x,y)$. The local magnetization is $z(\mathbf{x},t)=(n_1-n_2)/(n_1+n_2)$, while the local relative phase is $\phi(\mathbf{x},t)= \phi_1-\phi_2$. The total magnetization coincides with the spatial average of $z$, as $Z(t) =(1/{\cal A}) \int \! d\mathbf{x}\ z(\mathbf{x},t)$, and we define an average relative phase $\varphi$ such that $\cos \varphi(t) = (1/{\cal A}) \int \!d\mathbf{x} \cos \phi(\mathbf{x},t)$. 
The $c$-fields describe the low-lying macroscopically occupied modes of each components up to a cut-off energy $\epsilon_{\mathrm{cut}}=k_BT\ln(2)+\mu$, where the mean occupation number of single-particle modes is of order $1$. The projector $\hat{\mathcal{P}}$ confines the dynamics to the coherent fields below the cut-off, while higher-energy modes form a thermal reservoir that generates Gaussian noise satisfying the condition 
\begin{eqnarray}
\langle \eta_i(\mathbf{x}, t)\eta_j^*(\mathbf{x}', t') \rangle = 2\hbar\gamma k_BT\delta(\mathbf{x}-\mathbf{x}')\delta(t-t')\delta_{ij},
\label{eq:fluctuation-dissipation-theorem}
\end{eqnarray}
where $\langle \cdots \rangle$ denotes the averaging over different noise realizations. The same choice of energy cutoff has also been used previously to reproduce experimental results for single-component condensates in 2D configurations~\cite{miki_ota_18,Larcher2018}. This is consistent with earlier SPGPE-based studies of analogue false vacuum decay in ultracold atomic systems, where a cutoff of the form $\epsilon_{\mathrm{cut}} \sim k_B T$ (equivalently, $k_{\mathrm{cut}} = \sqrt{2T}$ in dimensionless units) has been adopted based on standard classical-field considerations ~\cite{Billam_2023,Billam_21,Billam_20}. Importantly, the exact value of the cutoff is not crucial, provided it remains within a physically reasonable range, as the results are largely insensitive to its precise choice~\cite{Liu_16, Billam_20}.
Following Refs.~\cite{liu_12,su_12, su_17}, we assume that the noise term is the same for both components. The dissipation parameter $\gamma$ couples the coherent region to the incoherent thermal reservoir. Its value affects the thermalization rate, but it does not alter the main properties of the final state after equilibration. Unless otherwise stated, we choose $\gamma=0.01$. Finally, for convenience, we introduce the temperature scale $T_s=|\kappa|n/k_B$, referred to as the effective spin temperature, to represent $T$ in dimensionless units.

It is worth stressing that, within SPGPE, results obtained from independent noise realizations may be regarded consistent with the representation of the outcomes of individual experimental runs (see, e.g.,~\cite{Sakmann_16,Sakmann_17,Olsen_17}). Due to the stochastic nature of the noise, each realization yields different dynamical trajectories, as it is also the case in repeated measurements. However, experiments cannot follow the temporal evolution of individual stochastic trajectories; a measurement at time $t$ simply captures the outcome of one realization within the underlying ensemble of trajectories \cite{zenesini_24}. 

\subsection{Initial state preparation}

The parameters entering Eq.~(\ref{eq:SGPE-eqn}) are chosen to emulate typical experimental conditions~\cite{gaunt_13,Ville_18}. For the computational box we choose $L_x = L_y=25\,\mu\text{m}$  and impose periodic boundary conditions, which are commonly used in numerical studies of vacuum decay in homogeneous trap geometries~\cite{Billam_2023,Billam_22}. However in real experiments, trap edges can act as preferential nucleation sites for bubbles~\cite{Billam_2023} and schemes have been proposed to mitigate such boundary effects~\cite{jenkins_25,K.brown_25}. Our simulations exhibit stochastic bubble formation in the bulk, with boundary effects absent by construction.
The chemical potential $\mu$, which is an input parameter in (\ref{eq:SGPE-eqn}), is chosen through Eq.~(\ref{chemical_potential}) so that the total number of atoms in the box is of the order of $10^4$. A natural energy scale is provided by $gn$, for which we have $gn/k_B = 8.5$~nK. For the mass we take $mg/\hbar^2 = 0.095$. The coherent coupling is chosen as $\Omega=0.03\,gn$, within the range accessible to current experiments~\cite{Farolfi_21,cominotti_22}. We finally take $g_{12}=1.2g$ and an initial detuning $\delta_i=0.5\Omega$. With these parameters, the effective spin temperature becomes $T_s=0.1 gn/k_B$. The highest absolute temperatures in our simulations are $\sim 7\,\mathrm{nK}$, which are lower than those typically reported in standard BEC experiments. This originates from the relatively low atom numbers (and hence low densities) chosen for our numerical simulations. However, the physically relevant parameter is the ratio between thermal and interaction energy; this implies that mixtures with higher density, within experimentally accessible regimes, will exhibit a similar behavior at larger temperatures. 
In our case, the ratio of thermal to interaction energy ($T/T_s$) lies in the range $5.5$–$8$. The ratio $\Omega/(|\kappa| n)$ remains in a comparable regime, and overall, the chosen parameters, while quantitatively distinct, lie within current experimental capabilities~\cite{zenesini_24,cominotti_25}.

Equilibrium configurations at a given temperature $T$ and detuning $\delta_i$ are obtained by numerically propagating Eq.~(\ref{eq:SGPE-eqn}) in real time from random initial $c$-fields until equilibration. In the SPGPE framework, the total number of atoms is not a direct input; rather, it evolves dynamically during simulations and stabilizes around a steady mean value. In our simulations we identify equilibrium states at a time when the total number of atoms saturates around a mean value close to $10^4$ and the fluctuations in the number of atoms around the respective time average is small~(typically within $1\%$). 

\begin{figure}
\centering
\includegraphics[width=1.0\linewidth]{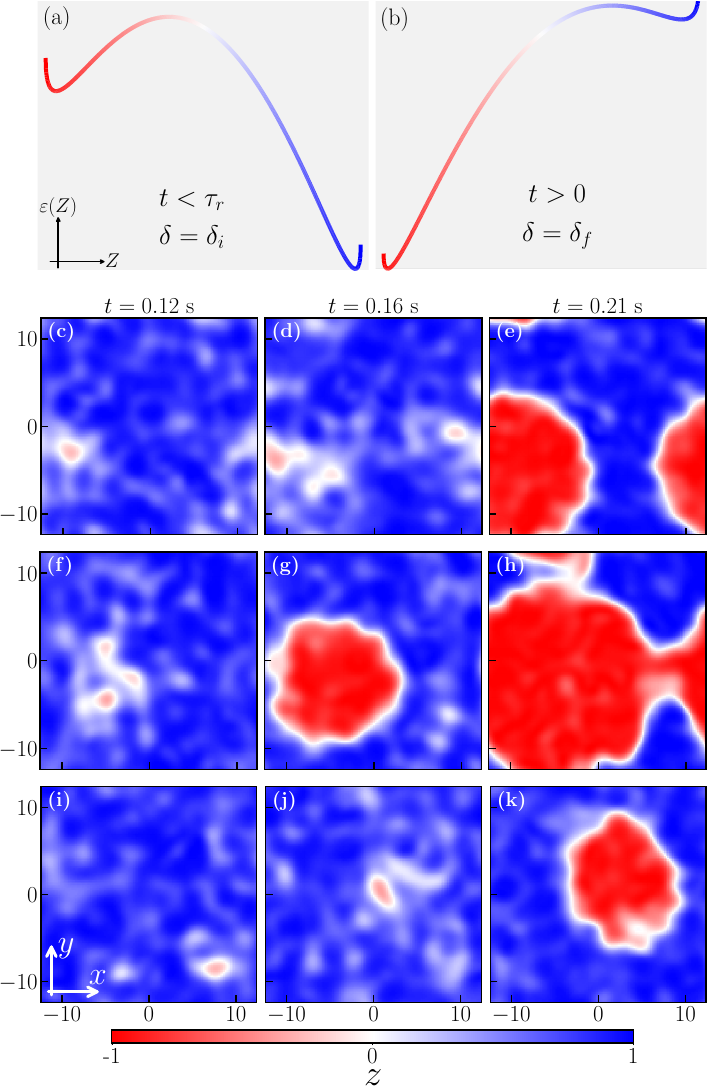}
\caption{Panels (a) and (b) show the energy landscapes, in arbitrary units, described by Eq.~(\ref{eq:energy_density}).
In (a), for an initial positive detuning $\delta_i = 0.5 \Omega$, the system is prepared at equilibrium near the global minimum ($Z \approx 1$) at temperature $T = 5.5T_s$. For $\tau_r< t < 0$, with $\tau_r=-1.6$~s, the detuning is linearly varied from the initial $\delta_i$ to a final negative value $\delta_f = -1.0 \Omega$, hence transferring the system into a metastable state, as in (b). The system subsequently decays into the true vacuum.
Panels (c)–(k) show typical snapshots of the local magnetization $z=(n_1-n_2)/(n_1+n_2)$ in the $x$–$y$ plane at three different times during the decay process, namely $120$, $160$ and $210$~ms. Each square box has length $25~\mu \rm{m}$ and periodic boundary conditions. Each row corresponds to a distinct stochastic noise realization. The figure shows bubbles of condensate $2$ (red) forming and growing in condensate $1$ (blue).} 

\label{fig:protocol}
\end{figure}

 After preparing the initial equilibrium state at the global minimum of the energy density (\ref{eq:energy_density}), the next task is to engineer the false vacuum state. To this end, we follow the experimental ramping protocol of Ref.~\cite{zenesini_24}, wherein the detuning is linearly varied from $\delta_i$ to the final value $\delta_f$ using $\delta(t)=\delta_i-(\delta_i-\delta_f)(1 - t/\tau_r)$, with the chemical potential $\mu$ being consistently corrected using Eqs.~(\ref{chemical_potential}) and~(\ref{eq:SGPE-eqn}). The time $\tau_r=-1.6~(s)$ sets the beginning of the ramp, which guides the system, almost adiabatically, towards the FV state, i.e., the local minimum of the effective energy landscape, as in Fig.~\ref{fig:protocol}~(b). 
 Although the fluctuation--dissipation theorem built into Eq.~(\ref{eq:SGPE-eqn}) should ensure that the initial state relaxes into the FV, in practice some trajectories may prematurely decay during the ramping process, especially at high temperatures. Given that our focus is on the decay of FV states, for the subsequent analysis, we restrict our attention to trajectories that have not decayed prematurely. In practice, we select the FV states for which ${Z}> 0.2$ at the end of the ramping stage $(t=0)$. We have checked that the results for the decay rates remain robust to the precise value of this threshold, provided that it corresponds to a positive magnetization.

\section{\label{sec:FVD}The decay of the false vacuum}

Once a metastable FV state is successfully prepared, we switch off the dissipation parameter $\gamma$ in (\ref{eq:SGPE-eqn}) and evolve the system under the resulting conservative projected Gross--Pitaevskii equation (PGPE). Setting $\gamma=0$ at this stage, ensures that the decay process is not affected by spurious processes caused by atom losses in the non-conservative SPGPE, while keeping the relevant effects of thermal fluctuations within the $c$-fields.  If $\gamma \neq 0$, the dynamics would acquire an additional time scale that hinders the intrinsic FVD mechanism, causing the observed decay to explicitly depend on $\gamma$. The same protocol has already been used in similar cases as, for instance, for the investigation of sound propagation in 2D condensates~\cite{Arko_21,Ville_18}.

The ensuing dynamics of the magnetization density reveals the decay of the FV towards the TV, which is characterized by the appearance and growth of bubbles of atoms of opposite polarization, as  shown in Fig.~\ref{fig:protocol}~(c)--(k). This is consistent with the semiclassical instanton picture, where decay occurs {\it via} the nucleation of resonant bubbles,
which are formed when the energy gain from the bulk is balanced by the surface energy of the bubble walls~\cite{Coleman_77,Callan_77}.
The nucleation probability per unit time and area is characterized by the decay rate $\Gamma$, which quantifies the transition of the false vacuum into a bubble configuration. Unlike single-particle tunneling, the decay rate in field theory exhibits a nontrivial dependence on the barrier parameters, which in our system are controlled by the externally tunable parameters $\delta$ and $\Omega$.
At finite temperatures, Linde’s extension of instanton theory shows that decay can proceed from thermally excited states, with the total rate obtained by summing over these contributions weighted by a Boltzmann factor~\cite{Linde_83,Ian_Affleck_81}. 
The decay rate takes the form 
\begin{equation}
\Gamma = A \, e^{-\beta E_c} \ ,
\label{Gamma}
\end{equation}
where $E_c$ is the critical instanton energy and $\beta = 1/k_B T$. The prefactor $A$ captures the fluctuations around the critical bubble and depends on the underlying system parameters; its temperature dependence is assumed to be weak compared to that of the exponential term. For a one-dimensional geometry, in very specific regimes, analytical results for $E_c$ and $A$ can be obtained \cite{zenesini_24,Garcia_26}. For our two-dimensional configuration there are no available results and in the following we treat them as fitting parameters.

Linde’s finite-temperature framework applies when thermal fluctuations dominate over quantum fluctuations ($T \gg T_s$) while still satisfying $k_B T \ll E_c$. For $k_B T \gg E_c$, the system simply rolls down the barrier without tunneling. 
In our numerical investigations we set the thermal energy $k_BT$ to be much lesser than the many-body energy associated with the barrier height described by Eq.~(\ref{eq:energy_density}). In particular, the ratio between thermal energy to barrier height is of the order of $10^{-2}$ in our simulations, placing it within reach of experimental conditions~\cite{zenesini_24}. It should, however, be emphasized that $E_c$ is not merely the energy of the barrier height, but the energy of the critical bubble configuration, for which there are no analytical predictions; its determination is beyond the scope of the current work.

In our numerical simulations, the dynamics can be analyzed in terms of both density and spin channels, which correspond to the in-phase and out-of phase modes, respectively. Within a hydrodynamic-like formalism, the system is usually described in terms of the local variables $n$, $z$ and $\phi$, where the density $n$ 
evolves according to the continuity equation, the magnetization $z$ is determined by the spin sector of the energy density, and the relative phase $\phi$ obeys 
an Euler-type equation~\cite{Alessio_22}.
In the presence of a weak Rabi coupling~($\Omega \ll \mu$), as in our case, and for $g_{12}\approx g$, the density channel is weakly affected by the perturbations in the spin channel. Consequently, the relevant variables of the system reduce to those of magnetization channel, i.e., magnetization and relative phase, which obey coupled Josephson equations~\cite{Alessio_22}. Since $z$ and $\phi$ are canonically conjugate variables, they are intrinsically coupled and cannot be treated independently.
The false vacuum state is therefore naturally described by a complex scalar field in which both degrees of freedom play an essential role in the process. In the following, we analyze the results of our numerical simulations, focusing first on the decay dynamics of the magnetization and then examining the associated dynamics of the relative phase.

\begin{figure}
    \centering
    \includegraphics[width=1.0\linewidth]{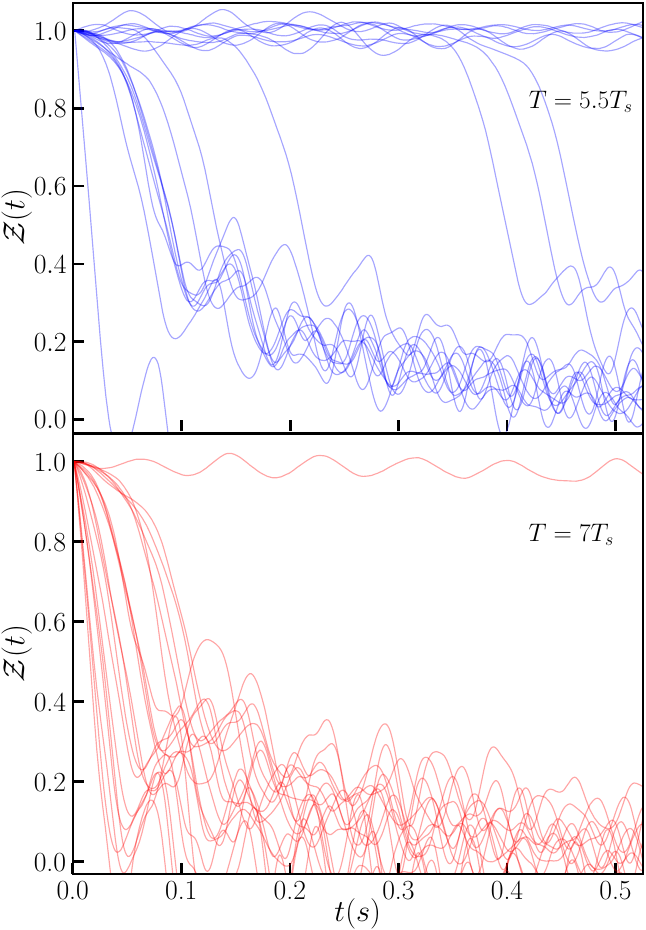}
    \caption{Representative trajectories of the rescaled magnetization ${\cal Z}(t)$ obtained by solving the PGPE from $20$ different stochastic realizations and two different temperatures. Here the detuning is $\delta_f = -\Omega$. The false vacuum decay occurs when ${\cal Z}(t)$ exhibits a sharp decrease, corresponding to the growth of bubbles with $Z \approx -1$ (true vacuum) in a medium with $Z \approx 1$ (false vacuum).  The variation in decay times reflects the intrinsic stochasticity of the process. At $T = 7T_s$, the trajectories exhibit a narrower spread and decay rapidly, whereas at $T = 5.5T_s$ the decay occurs slowly with greater variability.
}
    \label{fig:PGPE-rawdata}
\end{figure}

\begin{figure}
    \centering
    \includegraphics[width=1.0\linewidth]{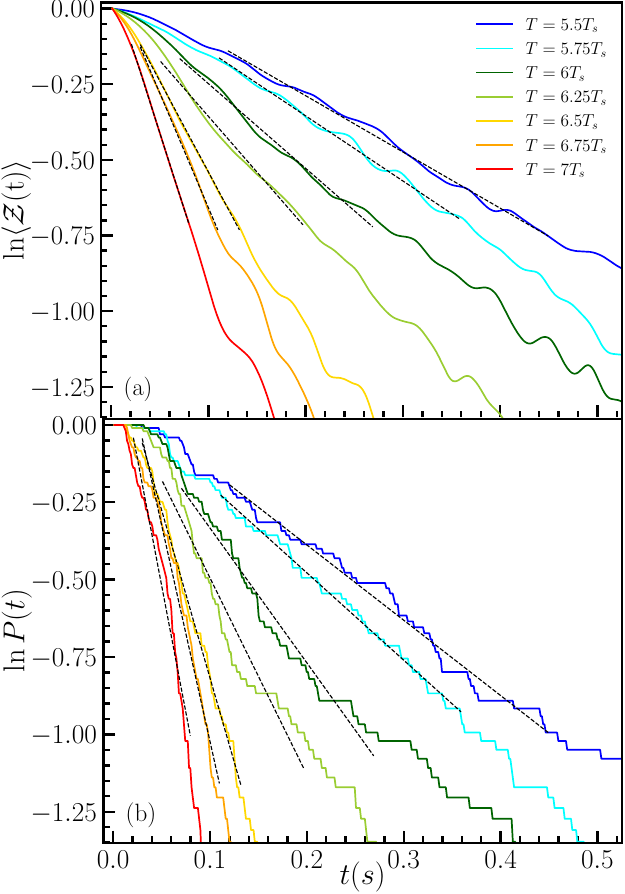}
    \caption{Logarithm of survival probability {\it vs.} time, at various temperatures and for  $\delta_f = -\Omega$. In (a) the survival probability is defined as the ensemble average of the rescaled magnetization, while in (b) is obtained by counting the number $P(t)$ of trajectories which are not yet decayed at time $t$.  All curves clearly exhibit exponential behavior, with both $\langle {\cal Z}(t) \rangle$ and $P(t)$ being proportional to $e^{-\Gamma t}$. The black dashed lines correspond to the fitting functions used to extract the decay rate $\Gamma$. For each temperature, the time interval used for the fit  starts approximately when the first growing bubble appears and ends when the true vacuum occupies about half of the box. }
    \label{fig:lnft_time}
\end{figure}

\subsection{Magnetization trajectories and survival probability}

To study the real-time dynamics of the decay process, we evolve the post-ramp state using the PGPE and monitor the behavior of the global magnetization $Z(t)$. In order to quantify the decay rate, we define the rescaled magnetization ${\cal Z}(t) = (1/2)\left(1 + Z(t)/Z(0)\right)$ and follow its time evolution for each noise realization at a given temperature~\cite{Lagnese_21}. 
This quantity is conveniently introduced in order to normalize the magnetization trajectories to a common initial value of unity, eliminating the spread in initial values with $Z>0.2$ at $t=0$.
Examples are shown in Fig.~\ref{fig:PGPE-rawdata} for $20$ noise realizations at $T = 5.5T_s$ (upper panel), and the same number at $7T_s$ (lower panel). We typically collect trajectories for  $\mathcal{N}=100$ different noise realizations at each temperature, and repeat it for different temperatures. The lowest temperature used in the simulations is selected to ensure sufficient decay 
statistics for estimating $\Gamma$, while the highest temperature~($T=8T_s$) is constrained by the prevalence 
of premature decays and is approximately $0.1T_{\rm{BKT}}$, where $T_{\rm{BKT}}$ is the critical temperature of the Berezinskii-Kosterlitz-Thouless transition for a single component Bose gas in the 2D geometry~\cite{prokofev_01}. 

As expected, most of the trajectories shown in Fig.~\ref{fig:PGPE-rawdata} exhibit a sharp decrease from ${\cal Z} = 1$ (false vacuum) towards ${\cal Z} \approx 0$ (true vacuum), but this decay occurs at different times depending on the initial noise. Also, the decay is systematically faster at higher temperature. From these trajectories, one can define a \textit{survival probability} in two different ways. The first consists of computing the ensemble average of the rescaled magnetization, $\langle {\cal Z}(t)\rangle$, over $\mathcal{N}$  noise realizations,  which is expected to follow an exponential law~\cite{Lagnese_21,zenesini_24}. The interpretation of this quantity as the \textit{survival probability} is not straightforward, since it involves an average over both surviving and decayed trajectories. Consequently, it also captures contributions from post bubble nucleation dynamics such as growth, expansion and  thermalization into the true vacuum. The second consists of counting the fraction $P(t)$ of trajectories that remain in the false vacuum for a time $t$, i.e., those with $Z(t) > 0$ (or equivalently ${\cal Z}(t) > 0.5$). This quantity has a direct probabilistic interpretation and is expected to exhibit exponential decay in time, thereby providing an alternative estimate of the survival probability~\cite{Pirvu_24}, that is less affected by the post-nucleation dynamics.
Our results for both protocols are shown in  Fig.~\ref{fig:lnft_time} in log-scale. In the long-time limit, all curves eventually decay, leading both $\langle {\cal Z}(t)\rangle$ and $P(t)$ to vanish. The stepwise behavior of $P(t)$, which is visible in Fig.~\ref{fig:lnft_time}(b), reflects the character of the counting procedure, which includes time intervals during which no decay events occur within our {\it ensemble} of trajectories. 

\begin{figure}
    \centering
    \includegraphics[width=1.0\linewidth]{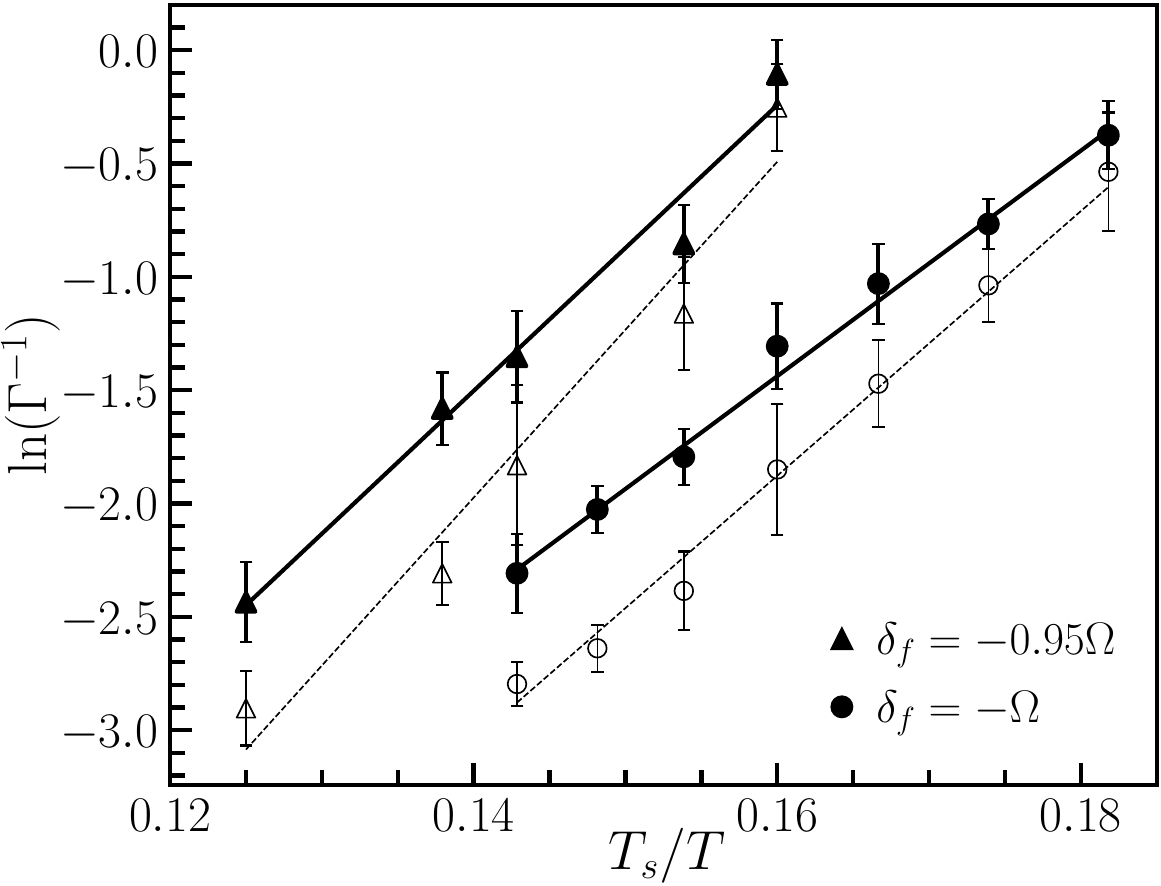}
         \caption{Temperature dependence of the decay rate $\Gamma$, extracted for $\delta_f=-\Omega$  and $\delta_f=-0.95\Omega$. The solid and empty markers represent the decay rate obtained from Fig.~\ref{fig:lnft_time}(a) and (b), respectively. The vertical error bars denote the statistical uncertainties in the estimated decay rates, obtained via the bootstrapping procedure~\cite{Billam_19,Bootstrapping}. The numerical data exhibit excellent agreement with the instanton prediction, $\Gamma \propto e^{-\beta E_c}$. For $\delta_f=-0.95\Omega$, the higher potential barrier leads to a reduced decay rate at any given temperature. 
         The straight lines are the fitted curves to the data represented through the markers.
         }
    \label{fig:Gamma_beta.pdf}
\end{figure}

\subsection{Decay rate}

The decay rate is extracted by fitting the curves of $\langle {\cal Z}(t)\rangle$ and $P(t)$ in a temporal window in which both exhibit a clear exponential decay~\cite{Billam_20,takacs_22} (see Fig.~\ref{fig:lnft_time}). In particular, the function $\ln \langle {\cal Z}(t)\rangle = \ln A_{{\cal Z}} - \Gamma t$ provides an excellent fit to the data for $\langle {\cal Z}(t)\rangle $ in the interval $[0.5, 0.9]$, with both $A_{{\cal Z}}$ and $\Gamma$ treated as fitting parameters. Within this temporal range, the cumulative bubble area remains below half of the system size, thereby avoiding significant finite-size effects. We repeat the fitting procedure also for $P(t)$ in the same temporal window, with the function $\ln P(t) = \ln A_{P} - \Gamma t$, again treating $A_{P}$ and $\Gamma$ as fitting parameters. 
Our numerical results for the decay rate $\Gamma$ are reported in Fig.~\ref{fig:Gamma_beta.pdf} for two values of the detuning $\delta_f$. The filled black markers are derived from $\langle {\cal Z}(t)\rangle $ (first protocol), while the empty markers from $P(t)$ (second protocol), with circles and triangles denoting the different detuning. The error bars accounts for the effects of ensemble averaging and are estimated {\it via} a bootstrapping procedure~\cite{Bootstrapping,Billam_19}.
In the figure, we actually plot the function $\ln(\Gamma^{-1})$ {\it vs.} $T_s/T$. The resulting linear dependence is consistent with the predictions of the instanton theory, expressed by Eq.~(\ref{Gamma}). The slope can therefore be identified as the critical instanton energy, $E_c/k_B$. From the first protocol~(thick solid black lines), for $\delta_f=-\Omega$, we obtain $E_c/k_B=(50 \pm 4$)~\rm{nK} and for $\delta_f=-0.95\Omega$, we find  $E_c/k_B=(63 \pm 6)~\rm{nK}$.  From the second protocol~(thin dashed lines), the corresponding slopes yield ~$(58 \pm 5)~\rm{nK}$  and $(74 \pm 7)~\rm{nK}$, respectively.  In all cases, uncertainties are estimated from the linear fit.
Although the numerical values of $\Gamma$ extracted from the two protocols are not identical, they both clearly show the expected exponential dependence on temperature. Furthermore, the linearity of the data indicates that, within the temperature values considered, $E_c$ remains temperature-independent. As expected, the slope, $E_c/k_B$ becomes larger for larger barrier height from both protocols. The positive slope reflects the enhancement of the decay rate with temperature, consistent with the thermal instanton theory.

\begin{figure}
    \centering
    \includegraphics[width=1.0\linewidth]{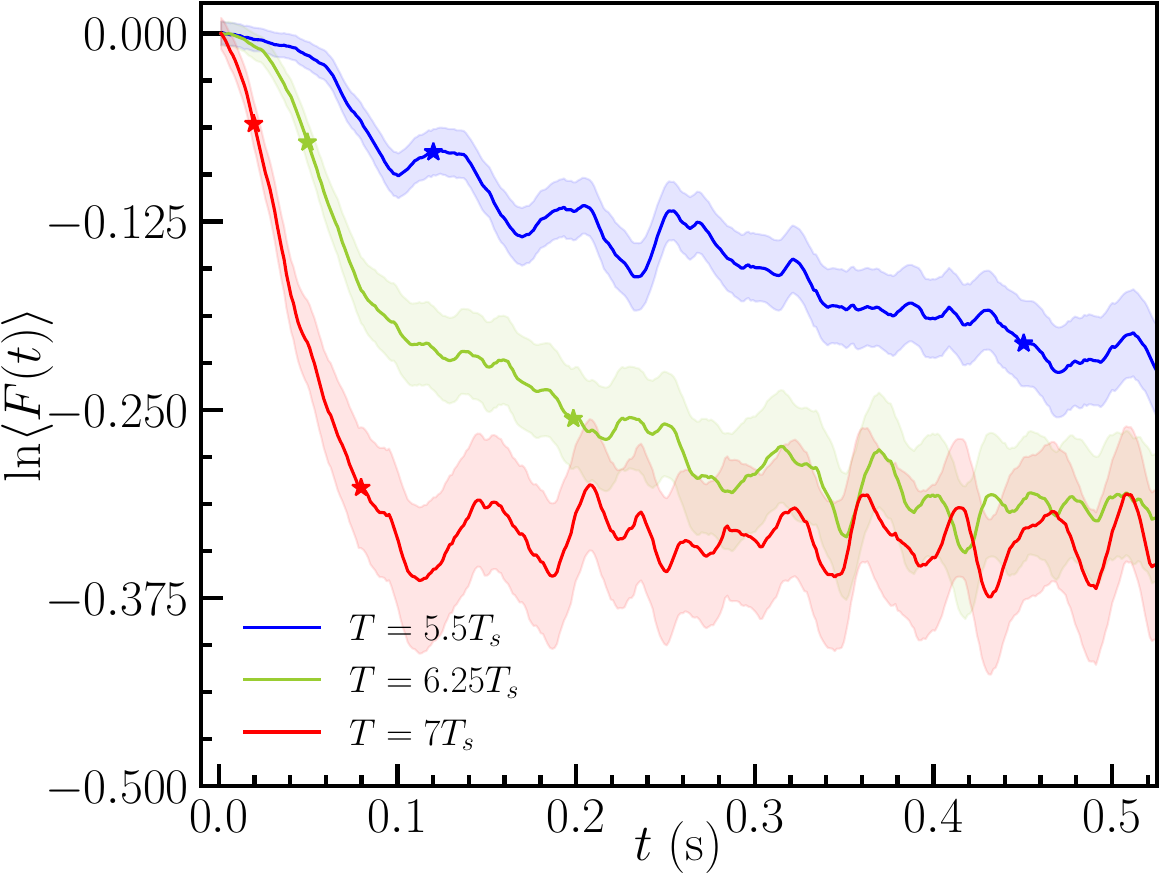}
         \caption{Relative phase dynamics at $\delta_f=-\Omega$ for three different temperatures. The lines represent the logarithm of the ensemble average of the rescaled $\cos \varphi$, where $\varphi$ is the relative phase and the shaded region denote the standard error bands for each temperature. For each line, a pair of markers delimits the time window used for the linear fit of the survival probability in Fig.~\ref{fig:lnft_time}}.
    \label{fig:fig6.pdf}
\end{figure}

\subsection{Relative phase dynamics}

 In PGPE simulations, we can also monitor the temporal evolution of the relative phase, as defined in Section~\ref{sec:SGPE}.
In non-equilibrium processes, such as false vacuum decay, the relative phase $\varphi$ can take any values in the range $[-\pi,\pi]$. Since the energy depends on $\cos\varphi$,
configurations with $\pm$ $\varphi$ are energetically identical, leading to the occurrence of solutions exhibiting both configurations. This
naturally motivates an estimate of the weight of the $\cos\varphi$
contribution in the energy landscape, Eq.~(\ref{eq:energy_density}). 
 In~\cite{zenesini_24,cominotti_25, Garcia_26}, it is assumed that the relative phase is locked during decay, so that the energy landscape is determined only by magnetization $Z$. Such an assumption is adopted to enable a tractable scalar-field treatment of instanton physics, in line with cosmological approaches~\cite{Devoto_22}. While this approximation successfully captures the main features of vacuum decay, it potentially limits a comprehensive understanding of the underlying system dynamics.
 
 Here we find that the relative phase actually evolves. To prove it, as we have already done for $Z(t)$ in Fig.~\ref{fig:lnft_time}, we introduce a rescaled function $F(t)=(1/2)\left(1 + \cos \varphi (t)/\cos \varphi(0)\right)$ and average it over $\mathcal{N}=100$ noise realizations.  Fig.~\ref{fig:fig6.pdf} shows the time evolution of $\ln \langle F(t) \rangle$ presented for three values of temperature. It can be seen that, over the same time interval during which the magnetization decays toward the true vacuum, the phase also changes, relaxing toward a constant value that exhibits only a weak dependence on temperature. Notably, the dominant contribution to this change occurs within the instanton window, i.e., the time interval used for the linear fitting of the survival probability in Fig.~\ref{fig:lnft_time}. In this interval, the variance of $\ln \langle F(t) \rangle$ (shaded region around the mean) is such that the curves for different temperatures remain clearly separated in the early stages of decay, demonstrating that the observed changes in the relative phase are statistically significant.
 For instance, for $T = 7T_s$, $\cos\varphi$ evolves from an initial value close to $-1$ and saturates around $-0.45$, corresponding to an approximate increase of $28\%$. This variation can significantly affect how the system crosses the barrier in the energy landscape. By inserting the ensemble averages of $Z(t)$ and $\cos \varphi(t)$ into (\ref{eq:energy_density}), we find that the energy values at the extremes of the time window used to extract the decay rates are comparable and that the energy has its maximum within the same interval. This suggests that the behavior of the relative phase is relevant for ensuring the conditions for resonant bubble nucleation. It also emphasizes the importance of using complex scalar fields in the description of coherently coupled Bose superfluids such that $Z$ and $\varphi$ are treated on equal footing, thus allowing a better understanding of the instanton physics.

\section{\label{sec:conclusions}Conclusions}

We have investigated finite-temperature false vacuum decay in coherently 
coupled two-dimensional homogeneous Bose–Einstein condensates, employing the 
Stochastic Projected Gross-Pitaevskii formalism to prepare thermal false-vacuum states and 
track their decay through global magnetization dynamics. Our analysis shows that the decay rate exhibits an exponential dependence on temperature, 
$\Gamma \propto e^{-\beta E_c}$, in agreement with thermal instanton theory. Furthermore, we elucidate the dynamics of the relative phase and suggest the need to develop an instanton theory describing a complex scalar field considering both magnetization and relative phase as relevant degrees of freedom, such as in coherently coupled Bose-Bose mixtures.

Our study places ultracold atoms as a viable platform to probe field-theoretic phenomena like vacuum decay, with the possibility of near-term experimental validation.
Given the high computational 
cost of large-scale SPGPE simulations, our present study is restricted to a given system size; a systematic exploration of barrier scaling and
finite-size effects remains an important direction for further study. Finally, while we have noted that false vacuum decay shares qualitative features with the Kramers escape problem~\cite{Berera_19}, in particular an exponential dependence of the decay rate in the thermally dominated regime, a field-theoretical analysis is needed to clarify their fundamental differences, and the development of such an analysis constitutes a promising avenue for subsequent research.

\begin{acknowledgments}
We thank Subhadeep Patra and Sunilkumar Venkateshappa for
several insightful discussions. Sivasankar gratefully acknowledges M.Obasho for constructive discussions on GPU-based implementation.
 A.Roy acknowledges the support of the Science and Engineering Research Board (SERB),
Department of Science and Technology, Government of India, under the project
SRG/2022/000057 and IIT Mandi seed-grant funds under the project IITM/SG/AR/87.
A.Roy acknowledges the National Supercomputing Mission (NSM) for providing
computing resources of PARAM Himalaya at IIT Mandi, which is implemented by
C-DAC and supported by the Ministry of Electronics and Information Technology
(MeitY) and Department of Science and Technology (DST), Government of India. F.Dalfovo and A.Recati thank the Provincia autonoma di Trento for support. A Recati acknowledges the Italian National Institute for Nuclear Physics (INFN) through the RELAQS project. This work is funded by the European Union under GA n$^\circ$101199451-QFIELBS. Views and opinions expressed are, however, those of the authors only and do not necessarily reflect those of the European Union or the European Research Council Executive Agency.  Neither the European Union nor the European Research Council Executive Agency can be held responsible for them.
 
\end{acknowledgments}

\bibliography{reference}

\begin{thebibliography}{72}%
\makeatletter
\providecommand \@ifxundefined [1]{%
 \@ifx{#1\undefined}
}%
\providecommand \@ifnum [1]{%
 \ifnum #1\expandafter \@firstoftwo
 \else \expandafter \@secondoftwo
 \fi
}%
\providecommand \@ifx [1]{%
 \ifx #1\expandafter \@firstoftwo
 \else \expandafter \@secondoftwo
 \fi
}%
\providecommand \natexlab [1]{#1}%
\providecommand \enquote  [1]{``#1''}%
\providecommand \bibnamefont  [1]{#1}%
\providecommand \bibfnamefont [1]{#1}%
\providecommand \citenamefont [1]{#1}%
\providecommand \href@noop [0]{\@secondoftwo}%
\providecommand \href [0]{\begingroup \@sanitize@url \@href}%
\providecommand \@href[1]{\@@startlink{#1}\@@href}%
\providecommand \@@href[1]{\endgroup#1\@@endlink}%
\providecommand \@sanitize@url [0]{\catcode `\\12\catcode `\$12\catcode
  `\&12\catcode `\#12\catcode `\^12\catcode `\_12\catcode `\%12\relax}%
\providecommand \@@startlink[1]{}%
\providecommand \@@endlink[0]{}%
\providecommand \url  [0]{\begingroup\@sanitize@url \@url }%
\providecommand \@url [1]{\endgroup\@href {#1}{\urlprefix }}%
\providecommand \urlprefix  [0]{URL }%
\providecommand \Eprint [0]{\href }%
\providecommand \doibase [0]{https://doi.org/}%
\providecommand \selectlanguage [0]{\@gobble}%
\providecommand \bibinfo  [0]{\@secondoftwo}%
\providecommand \bibfield  [0]{\@secondoftwo}%
\providecommand \translation [1]{[#1]}%
\providecommand \BibitemOpen [0]{}%
\providecommand \bibitemStop [0]{}%
\providecommand \bibitemNoStop [0]{.\EOS\space}%
\providecommand \EOS [0]{\spacefactor3000\relax}%
\providecommand \BibitemShut  [1]{\csname bibitem#1\endcsname}%
\let\auto@bib@innerbib\@empty
\bibitem [{\citenamefont {Coleman}(1977)}]{Coleman_77}%
  \BibitemOpen
  \bibfield  {author} {\bibinfo {author} {\bibfnamefont {S.}~\bibnamefont
  {Coleman}},\ }\bibfield  {title} {\bibinfo {title} {Fate of the false vacuum:
  Semiclassical theory},\ }\href {https://doi.org/10.1103/PhysRevD.15.2929}
  {\bibfield  {journal} {\bibinfo  {journal} {Phys. Rev. D}\ }\textbf {\bibinfo
  {volume} {15}},\ \bibinfo {pages} {2929} (\bibinfo {year}
  {1977})}\BibitemShut {NoStop}%
\bibitem [{\citenamefont {Callan}\ and\ \citenamefont
  {Coleman}(1977)}]{Callan_77}%
  \BibitemOpen
  \bibfield  {author} {\bibinfo {author} {\bibfnamefont {C.~G.}\ \bibnamefont
  {Callan}}\ and\ \bibinfo {author} {\bibfnamefont {S.}~\bibnamefont
  {Coleman}},\ }\bibfield  {title} {\bibinfo {title} {Fate of the false vacuum.
  ii. first quantum corrections},\ }\href
  {https://doi.org/10.1103/PhysRevD.16.1762} {\bibfield  {journal} {\bibinfo
  {journal} {Phys. Rev. D}\ }\textbf {\bibinfo {volume} {16}},\ \bibinfo
  {pages} {1762} (\bibinfo {year} {1977})}\BibitemShut {NoStop}%
\bibitem [{\citenamefont {Kobzarev}\ \emph {et~al.}(1974)\citenamefont
  {Kobzarev}, \citenamefont {Okun},\ and\ \citenamefont
  {Voloshin}}]{Kobzarev_74}%
  \BibitemOpen
  \bibfield  {author} {\bibinfo {author} {\bibfnamefont {I.~Y.}\ \bibnamefont
  {Kobzarev}}, \bibinfo {author} {\bibfnamefont {L.~B.}\ \bibnamefont {Okun}},\
  and\ \bibinfo {author} {\bibfnamefont {M.~B.}\ \bibnamefont {Voloshin}},\
  }\bibfield  {title} {\bibinfo {title} {Bubbles in {M}etastable {V}acuum},\
  }\href@noop {} {\bibfield  {journal} {\bibinfo  {journal} {Yad. Fiz.}\
  }\textbf {\bibinfo {volume} {20}},\ \bibinfo {pages} {1229} (\bibinfo {year}
  {1974})}\BibitemShut {NoStop}%
\bibitem [{\citenamefont {Coleman}\ and\ \citenamefont
  {De~Luccia}(1980)}]{Coleman_80}%
  \BibitemOpen
  \bibfield  {author} {\bibinfo {author} {\bibfnamefont {S.}~\bibnamefont
  {Coleman}}\ and\ \bibinfo {author} {\bibfnamefont {F.}~\bibnamefont
  {De~Luccia}},\ }\bibfield  {title} {\bibinfo {title} {Gravitational effects
  on and of vacuum decay},\ }\href {https://doi.org/10.1103/PhysRevD.21.3305}
  {\bibfield  {journal} {\bibinfo  {journal} {Phys. Rev. D}\ }\textbf {\bibinfo
  {volume} {21}},\ \bibinfo {pages} {3305} (\bibinfo {year}
  {1980})}\BibitemShut {NoStop}%
\bibitem [{\citenamefont {Hindmarsh}\ \emph {et~al.}(2021)\citenamefont
  {Hindmarsh}, \citenamefont {Lüben}, \citenamefont {Lumma},\ and\
  \citenamefont {Pauly}}]{M.hindmarsh_21}%
  \BibitemOpen
  \bibfield  {author} {\bibinfo {author} {\bibfnamefont {M.}~\bibnamefont
  {Hindmarsh}}, \bibinfo {author} {\bibfnamefont {M.}~\bibnamefont {Lüben}},
  \bibinfo {author} {\bibfnamefont {J.}~\bibnamefont {Lumma}},\ and\ \bibinfo
  {author} {\bibfnamefont {M.}~\bibnamefont {Pauly}},\ }\bibfield  {title}
  {\bibinfo {title} {Phase transitions in the early universe},\ }\href
  {https://doi.org/10.21468/SciPostPhysLectNotes.24} {\bibfield  {journal}
  {\bibinfo  {journal} {SciPost Phys. Lect. Notes}\ ,\ \bibinfo {pages} {24}}
  (\bibinfo {year} {2021})}\BibitemShut {NoStop}%
\bibitem [{\citenamefont {Burda}\ \emph {et~al.}(2015)\citenamefont {Burda},
  \citenamefont {Gregory},\ and\ \citenamefont {Moss}}]{Burda_15}%
  \BibitemOpen
  \bibfield  {author} {\bibinfo {author} {\bibfnamefont {P.}~\bibnamefont
  {Burda}}, \bibinfo {author} {\bibfnamefont {R.}~\bibnamefont {Gregory}},\
  and\ \bibinfo {author} {\bibfnamefont {I.~G.}\ \bibnamefont {Moss}},\
  }\bibfield  {title} {\bibinfo {title} {Gravity and the stability of the
  {H}iggs vacuum},\ }\href {https://doi.org/10.1103/PhysRevLett.115.071303}
  {\bibfield  {journal} {\bibinfo  {journal} {Phys. Rev. Lett.}\ }\textbf
  {\bibinfo {volume} {115}},\ \bibinfo {pages} {071303} (\bibinfo {year}
  {2015})}\BibitemShut {NoStop}%
\bibitem [{\citenamefont {Stone}(1977)}]{MSTONE_77}%
  \BibitemOpen
  \bibfield  {author} {\bibinfo {author} {\bibfnamefont {M.}~\bibnamefont
  {Stone}},\ }\bibfield  {title} {\bibinfo {title} {Semiclassical methods for
  unstable states},\ }\href {https://doi.org/10.1016/0370-2693(77)90099-5}
  {\bibfield  {journal} {\bibinfo  {journal} {Phys. Lett. B}\ }\textbf
  {\bibinfo {volume} {67}},\ \bibinfo {pages} {186} (\bibinfo {year}
  {1977})}\BibitemShut {NoStop}%
\bibitem [{\citenamefont {Shaposhnikov}(1987)}]{SHAPOSHNIKOV_87}%
  \BibitemOpen
  \bibfield  {author} {\bibinfo {author} {\bibfnamefont {M.~E.}\ \bibnamefont
  {Shaposhnikov}},\ }\bibfield  {title} {\bibinfo {title} {Baryon asymmetry of
  the universe in standard electroweak theory},\ }\href
  {https://doi.org/10.1016/0550-3213(87)90127-1} {\bibfield  {journal}
  {\bibinfo  {journal} {Nucl. Phys. B}\ }\textbf {\bibinfo {volume} {287}},\
  \bibinfo {pages} {757} (\bibinfo {year} {1987})}\BibitemShut {NoStop}%
\bibitem [{\citenamefont {Batini}\ \emph {et~al.}(2024)\citenamefont {Batini},
  \citenamefont {Chatrchyan},\ and\ \citenamefont {Berges}}]{Batini_24}%
  \BibitemOpen
  \bibfield  {author} {\bibinfo {author} {\bibfnamefont {L.}~\bibnamefont
  {Batini}}, \bibinfo {author} {\bibfnamefont {A.}~\bibnamefont {Chatrchyan}},\
  and\ \bibinfo {author} {\bibfnamefont {J.}~\bibnamefont {Berges}},\
  }\bibfield  {title} {\bibinfo {title} {Real-time dynamics of false vacuum
  decay},\ }\href {https://doi.org/10.1103/PhysRevD.109.023502} {\bibfield
  {journal} {\bibinfo  {journal} {Phys. Rev. D}\ }\textbf {\bibinfo {volume}
  {109}},\ \bibinfo {pages} {023502} (\bibinfo {year} {2024})}\BibitemShut
  {NoStop}%
\bibitem [{\citenamefont {{H}irvonen}\ and\ \citenamefont
  {{G}ould}(2026)}]{Hirvonen_25}%
  \BibitemOpen
  \bibfield  {author} {\bibinfo {author} {\bibfnamefont {J.}~\bibnamefont
  {{H}irvonen}}\ and\ \bibinfo {author} {\bibfnamefont {O.}~\bibnamefont
  {{G}ould}},\ }\bibfield  {title} {\bibinfo {title} {{L}anger's {N}ucleation
  {R}ate {R}eproduced on the {L}attice},\ }\href
  {https://doi.org/10.1103/c75g-xbw4} {\bibfield  {journal} {\bibinfo
  {journal} {Phys. Rev. Lett.}\ }\textbf {\bibinfo {volume} {136}},\ \bibinfo
  {pages} {081601} (\bibinfo {year} {2026})}\BibitemShut {NoStop}%
\bibitem [{\citenamefont {Baldwin}\ \emph {et~al.}(2011)\citenamefont
  {Baldwin}, \citenamefont {Knowles}, \citenamefont {Tartaglia}, \citenamefont
  {Fitzpatrick}, \citenamefont {Devlin}, \citenamefont {Shammas}, \citenamefont
  {Waudby}, \citenamefont {Mossuto}, \citenamefont {Meehan}, \citenamefont
  {Gras}, \citenamefont {Christodoulou}, \citenamefont {Anthony-Cahill},
  \citenamefont {Barker}, \citenamefont {Vendruscolo},\ and\ \citenamefont
  {Dobson}}]{Baldwin_11}%
  \BibitemOpen
  \bibfield  {author} {\bibinfo {author} {\bibfnamefont {A.~J.}\ \bibnamefont
  {Baldwin}}, \bibinfo {author} {\bibfnamefont {T.~P.~J.}\ \bibnamefont
  {Knowles}}, \bibinfo {author} {\bibfnamefont {G.~G.}\ \bibnamefont
  {Tartaglia}}, \bibinfo {author} {\bibfnamefont {A.~W.}\ \bibnamefont
  {Fitzpatrick}}, \bibinfo {author} {\bibfnamefont {G.~L.}\ \bibnamefont
  {Devlin}}, \bibinfo {author} {\bibfnamefont {S.~L.}\ \bibnamefont {Shammas}},
  \bibinfo {author} {\bibfnamefont {C.~A.}\ \bibnamefont {Waudby}}, \bibinfo
  {author} {\bibfnamefont {M.~F.}\ \bibnamefont {Mossuto}}, \bibinfo {author}
  {\bibfnamefont {S.}~\bibnamefont {Meehan}}, \bibinfo {author} {\bibfnamefont
  {S.~L.}\ \bibnamefont {Gras}}, \bibinfo {author} {\bibfnamefont
  {J.}~\bibnamefont {Christodoulou}}, \bibinfo {author} {\bibfnamefont {S.~J.}\
  \bibnamefont {Anthony-Cahill}}, \bibinfo {author} {\bibfnamefont {P.~D.}\
  \bibnamefont {Barker}}, \bibinfo {author} {\bibfnamefont {M.}~\bibnamefont
  {Vendruscolo}},\ and\ \bibinfo {author} {\bibfnamefont {C.~M.}\ \bibnamefont
  {Dobson}},\ }\bibfield  {title} {\bibinfo {title} {Metastability of native
  proteins and the phenomenon of amyloid formation},\ }\href
  {https://doi.org/10.1021/ja2017703} {\bibfield  {journal} {\bibinfo
  {journal} {J. Am. Chem. Soc.}\ }\textbf {\bibinfo {volume} {133}},\ \bibinfo
  {pages} {14160} (\bibinfo {year} {2011})}\BibitemShut {NoStop}%
\bibitem [{\citenamefont {Ghosh}\ and\ \citenamefont
  {Ranjan}(2020)}]{Ghosh_20}%
  \BibitemOpen
  \bibfield  {author} {\bibinfo {author} {\bibfnamefont {D.~K.}\ \bibnamefont
  {Ghosh}}\ and\ \bibinfo {author} {\bibfnamefont {A.}~\bibnamefont {Ranjan}},\
  }\bibfield  {title} {\bibinfo {title} {The metastable states of proteins},\
  }\href {https://doi.org/https://doi.org/10.1002/pro.3859} {\bibfield
  {journal} {\bibinfo  {journal} {Protein Sci.}\ }\textbf {\bibinfo {volume}
  {29}},\ \bibinfo {pages} {1559} (\bibinfo {year} {2020})}\BibitemShut
  {NoStop}%
\bibitem [{\citenamefont {Oxtoby}(1992)}]{DWOxtoby_92}%
  \BibitemOpen
  \bibfield  {author} {\bibinfo {author} {\bibfnamefont {D.~W.}\ \bibnamefont
  {Oxtoby}},\ }\bibfield  {title} {\bibinfo {title} {Homogeneous nucleation:
  theory and experiment},\ }\href {https://doi.org/10.1088/0953-8984/4/38/001}
  {\bibfield  {journal} {\bibinfo  {journal} {J. Phys.: Condens. Matter}\
  }\textbf {\bibinfo {volume} {4}},\ \bibinfo {pages} {7627} (\bibinfo {year}
  {1992})}\BibitemShut {NoStop}%
\bibitem [{\citenamefont {Debenedetti}\ and\ \citenamefont
  {Stillinger}(2001)}]{Debenedetti_01}%
  \BibitemOpen
  \bibfield  {author} {\bibinfo {author} {\bibfnamefont {P.~G.}\ \bibnamefont
  {Debenedetti}}\ and\ \bibinfo {author} {\bibfnamefont {F.~H.}\ \bibnamefont
  {Stillinger}},\ }\bibfield  {title} {\bibinfo {title} {Supercooled liquids
  and the glass transition},\ }\href {https://doi.org/10.1038/35065704}
  {\bibfield  {journal} {\bibinfo  {journal} {Nature}\ }\textbf {\bibinfo
  {volume} {410}},\ \bibinfo {pages} {259} (\bibinfo {year}
  {2001})}\BibitemShut {NoStop}%
\bibitem [{\citenamefont {Coldea}\ \emph {et~al.}(2010)\citenamefont {Coldea},
  \citenamefont {Tennant}, \citenamefont {Wheeler}, \citenamefont {Wawrzynska},
  \citenamefont {Prabhakaran}, \citenamefont {Telling}, \citenamefont
  {Habicht}, \citenamefont {Smeibidl},\ and\ \citenamefont
  {Kiefer}}]{R.Coldea_10}%
  \BibitemOpen
  \bibfield  {author} {\bibinfo {author} {\bibfnamefont {R.}~\bibnamefont
  {Coldea}}, \bibinfo {author} {\bibfnamefont {D.~A.}\ \bibnamefont {Tennant}},
  \bibinfo {author} {\bibfnamefont {E.~M.}\ \bibnamefont {Wheeler}}, \bibinfo
  {author} {\bibfnamefont {E.}~\bibnamefont {Wawrzynska}}, \bibinfo {author}
  {\bibfnamefont {D.}~\bibnamefont {Prabhakaran}}, \bibinfo {author}
  {\bibfnamefont {M.}~\bibnamefont {Telling}}, \bibinfo {author} {\bibfnamefont
  {K.}~\bibnamefont {Habicht}}, \bibinfo {author} {\bibfnamefont
  {P.}~\bibnamefont {Smeibidl}},\ and\ \bibinfo {author} {\bibfnamefont
  {K.}~\bibnamefont {Kiefer}},\ }\bibfield  {title} {\bibinfo {title} {Quantum
  criticality in an {I}sing chain: Experimental evidence for emergent ${E}_{8}$
  symmetry},\ }\href {https://doi.org/10.1126/science.1180085} {\bibfield
  {journal} {\bibinfo  {journal} {Science}\ }\textbf {\bibinfo {volume}
  {327}},\ \bibinfo {pages} {177} (\bibinfo {year} {2010})}\BibitemShut
  {NoStop}%
\bibitem [{\citenamefont {Rutkevich}(1999)}]{Rutkevich_99}%
  \BibitemOpen
  \bibfield  {author} {\bibinfo {author} {\bibfnamefont {S.~B.}\ \bibnamefont
  {Rutkevich}},\ }\bibfield  {title} {\bibinfo {title} {Decay of the metastable
  phase in $d=1$ and $d=2$ {I}sing models},\ }\href
  {https://doi.org/10.1103/PhysRevB.60.14525} {\bibfield  {journal} {\bibinfo
  {journal} {Phys. Rev. B}\ }\textbf {\bibinfo {volume} {60}},\ \bibinfo
  {pages} {14525} (\bibinfo {year} {1999})}\BibitemShut {NoStop}%
\bibitem [{\citenamefont {Sinha}\ \emph {et~al.}(2021)\citenamefont {Sinha},
  \citenamefont {Chanda},\ and\ \citenamefont {Dziarmaga}}]{A.sinha_21}%
  \BibitemOpen
  \bibfield  {author} {\bibinfo {author} {\bibfnamefont {A.}~\bibnamefont
  {Sinha}}, \bibinfo {author} {\bibfnamefont {T.}~\bibnamefont {Chanda}},\ and\
  \bibinfo {author} {\bibfnamefont {J.}~\bibnamefont {Dziarmaga}},\ }\bibfield
  {title} {\bibinfo {title} {Nonadiabatic dynamics across a first-order quantum
  phase transition: Quantized bubble nucleation},\ }\href
  {https://doi.org/10.1103/PhysRevB.103.L220302} {\bibfield  {journal}
  {\bibinfo  {journal} {Phys. Rev. B}\ }\textbf {\bibinfo {volume} {103}},\
  \bibinfo {pages} {L220302} (\bibinfo {year} {2021})}\BibitemShut {NoStop}%
\bibitem [{\citenamefont {Lagnese}\ \emph {et~al.}(2024)\citenamefont
  {Lagnese}, \citenamefont {Surace}, \citenamefont {Morampudi},\ and\
  \citenamefont {Wilczek}}]{Lagnese_24}%
  \BibitemOpen
  \bibfield  {author} {\bibinfo {author} {\bibfnamefont {G.}~\bibnamefont
  {Lagnese}}, \bibinfo {author} {\bibfnamefont {F.~M.}\ \bibnamefont {Surace}},
  \bibinfo {author} {\bibfnamefont {S.}~\bibnamefont {Morampudi}},\ and\
  \bibinfo {author} {\bibfnamefont {F.}~\bibnamefont {Wilczek}},\ }\bibfield
  {title} {\bibinfo {title} {Detecting a long-lived false vacuum with quantum
  quenches},\ }\href {https://doi.org/10.1103/PhysRevLett.133.240402}
  {\bibfield  {journal} {\bibinfo  {journal} {Phys. Rev. Lett.}\ }\textbf
  {\bibinfo {volume} {133}},\ \bibinfo {pages} {240402} (\bibinfo {year}
  {2024})}\BibitemShut {NoStop}%
\bibitem [{\citenamefont {Johansen}\ \emph {et~al.}(2025)\citenamefont
  {Johansen}, \citenamefont {Recati}, \citenamefont {Carusotto},\ and\
  \citenamefont {Biella}}]{johansen_25}%
  \BibitemOpen
  \bibfield  {author} {\bibinfo {author} {\bibfnamefont {C.}~\bibnamefont
  {Johansen}}, \bibinfo {author} {\bibfnamefont {A.}~\bibnamefont {Recati}},
  \bibinfo {author} {\bibfnamefont {I.}~\bibnamefont {Carusotto}},\ and\
  \bibinfo {author} {\bibfnamefont {A.}~\bibnamefont {Biella}},\ }\href@noop {}
  {\bibinfo {title} {Many-body theory of false vacuum decay in quantum spin
  chains}} (\bibinfo {year} {2025}),\ \Eprint
  {https://arxiv.org/abs/2508.13780} {arXiv:2508.13780} \BibitemShut {NoStop}%
\bibitem [{\citenamefont {Pavešić}\ \emph {et~al.}(2025)\citenamefont
  {Pavešić}, \citenamefont {Di~Liberto},\ and\ \citenamefont
  {Montangero}}]{luka_pavesic_25}%
  \BibitemOpen
  \bibfield  {author} {\bibinfo {author} {\bibfnamefont {L.}~\bibnamefont
  {Pavešić}}, \bibinfo {author} {\bibfnamefont {M.}~\bibnamefont
  {Di~Liberto}},\ and\ \bibinfo {author} {\bibfnamefont {S.}~\bibnamefont
  {Montangero}},\ }\href@noop {} {\bibinfo {title} {Scattering and induced
  false vacuum decay in the two-dimensional quantum {I}sing model}} (\bibinfo
  {year} {2025}),\ \Eprint {https://arxiv.org/abs/2509.02702}
  {arXiv:2509.02702} \BibitemShut {NoStop}%
\bibitem [{\citenamefont {Maertens}\ \emph {et~al.}(2025)\citenamefont
  {Maertens}, \citenamefont {Haegeman},\ and\ \citenamefont
  {Van~Acoleyen}}]{daan_maertens_25}%
  \BibitemOpen
  \bibfield  {author} {\bibinfo {author} {\bibfnamefont {D.}~\bibnamefont
  {Maertens}}, \bibinfo {author} {\bibfnamefont {J.}~\bibnamefont {Haegeman}},\
  and\ \bibinfo {author} {\bibfnamefont {K.}~\bibnamefont {Van~Acoleyen}},\
  }\href@noop {} {\bibinfo {title} {Real-time bubble nucleation and growth for
  false vacuum decay on the lattice}} (\bibinfo {year} {2025}),\ \Eprint
  {https://arxiv.org/abs/2508.13645} {arXiv:2508.13645} \BibitemShut {NoStop}%
\bibitem [{\citenamefont {Borla}\ \emph {et~al.}(2026)\citenamefont {Borla},
  \citenamefont {Lazarides}, \citenamefont {Groß},\ and\ \citenamefont
  {Halimeh}}]{Borla_26}%
  \BibitemOpen
  \bibfield  {author} {\bibinfo {author} {\bibfnamefont {U.}~\bibnamefont
  {Borla}}, \bibinfo {author} {\bibfnamefont {A.}~\bibnamefont {Lazarides}},
  \bibinfo {author} {\bibfnamefont {C.}~\bibnamefont {Groß}},\ and\ \bibinfo
  {author} {\bibfnamefont {J.~C.}\ \bibnamefont {Halimeh}},\ }\href@noop {}
  {\bibinfo {title} {{M}icroscopic {D}ynamics of {F}alse {V}acuum {D}ecay in
  the $2+1${D} {Q}uantum {I}sing {M}odel}} (\bibinfo {year} {2026}),\ \Eprint
  {https://arxiv.org/abs/2601.04305} {arXiv:2601.04305} \BibitemShut {NoStop}%
\bibitem [{\citenamefont {Abel}\ and\ \citenamefont
  {Spannowsky}(2021)}]{Abel_21}%
  \BibitemOpen
  \bibfield  {author} {\bibinfo {author} {\bibfnamefont {S.}~\bibnamefont
  {Abel}}\ and\ \bibinfo {author} {\bibfnamefont {M.}~\bibnamefont
  {Spannowsky}},\ }\bibfield  {title} {\bibinfo {title}
  {Quantum-field-theoretic simulation platform for observing the fate of the
  false vacuum},\ }\href {https://doi.org/10.1103/PRXQuantum.2.010349}
  {\bibfield  {journal} {\bibinfo  {journal} {PRX Quantum}\ }\textbf {\bibinfo
  {volume} {2}},\ \bibinfo {pages} {010349} (\bibinfo {year}
  {2021})}\BibitemShut {NoStop}%
\bibitem [{\citenamefont {Vodeb}\ \emph {et~al.}(2025)\citenamefont {Vodeb},
  \citenamefont {Desaules}, \citenamefont {Hallam}, \citenamefont {Rava},
  \citenamefont {Humar}, \citenamefont {Willsch}, \citenamefont {Jin},
  \citenamefont {Willsch}, \citenamefont {Michielsen},\ and\ \citenamefont
  {Papi{\'{c}}}}]{Vodeb_25}%
  \BibitemOpen
  \bibfield  {author} {\bibinfo {author} {\bibfnamefont {J.}~\bibnamefont
  {Vodeb}}, \bibinfo {author} {\bibfnamefont {J.-Y.}\ \bibnamefont {Desaules}},
  \bibinfo {author} {\bibfnamefont {A.}~\bibnamefont {Hallam}}, \bibinfo
  {author} {\bibfnamefont {A.}~\bibnamefont {Rava}}, \bibinfo {author}
  {\bibfnamefont {G.}~\bibnamefont {Humar}}, \bibinfo {author} {\bibfnamefont
  {D.}~\bibnamefont {Willsch}}, \bibinfo {author} {\bibfnamefont
  {F.}~\bibnamefont {Jin}}, \bibinfo {author} {\bibfnamefont {M.}~\bibnamefont
  {Willsch}}, \bibinfo {author} {\bibfnamefont {K.}~\bibnamefont
  {Michielsen}},\ and\ \bibinfo {author} {\bibfnamefont {Z.}~\bibnamefont
  {Papi{\'{c}}}},\ }\bibfield  {title} {\bibinfo {title} {Stirring the false
  vacuum via interacting quantized bubbles on a 5,564-qubit quantum annealer},\
  }\href {https://doi.org/10.1038/s41567-024-02765-w} {\bibfield  {journal}
  {\bibinfo  {journal} {Nat. Phys.}\ }\textbf {\bibinfo {volume} {21}},\
  \bibinfo {pages} {386} (\bibinfo {year} {2025})}\BibitemShut {NoStop}%
\bibitem [{\citenamefont {Fialko}\ \emph {et~al.}(2017)\citenamefont {Fialko},
  \citenamefont {Opanchuk}, \citenamefont {Sidorov}, \citenamefont
  {{D}rummond},\ and\ \citenamefont {{B}rand}}]{Fialko_17}%
  \BibitemOpen
  \bibfield  {author} {\bibinfo {author} {\bibfnamefont {O.}~\bibnamefont
  {Fialko}}, \bibinfo {author} {\bibfnamefont {B.}~\bibnamefont {Opanchuk}},
  \bibinfo {author} {\bibfnamefont {A.~I.}\ \bibnamefont {Sidorov}}, \bibinfo
  {author} {\bibfnamefont {P.~D.}\ \bibnamefont {{D}rummond}},\ and\ \bibinfo
  {author} {\bibfnamefont {J.}~\bibnamefont {{B}rand}},\ }\bibfield  {title}
  {\bibinfo {title} {The universe on a table top: Engineering quantum decay of
  a relativistic scalar field from a metastable vacuum},\ }\href
  {https://doi.org/10.1088/1361-6455/50/2/024003} {\bibfield  {journal}
  {\bibinfo  {journal} {J. Phys. B}\ }\textbf {\bibinfo {volume} {50}},\
  \bibinfo {pages} {024003} (\bibinfo {year} {2017})}\BibitemShut {NoStop}%
\bibitem [{\citenamefont {Braden}\ \emph {et~al.}(2018)\citenamefont {Braden},
  \citenamefont {Johnson}, \citenamefont {Peiris},\ and\ \citenamefont
  {Weinfurtner}}]{braden_18}%
  \BibitemOpen
  \bibfield  {author} {\bibinfo {author} {\bibfnamefont {J.}~\bibnamefont
  {Braden}}, \bibinfo {author} {\bibfnamefont {M.~C.}\ \bibnamefont {Johnson}},
  \bibinfo {author} {\bibfnamefont {H.~V.}\ \bibnamefont {Peiris}},\ and\
  \bibinfo {author} {\bibfnamefont {S.}~\bibnamefont {Weinfurtner}},\
  }\bibfield  {title} {\bibinfo {title} {Towards the cold atom analog false
  vacuum},\ }\href {https://doi.org/10.1007/JHEP07(2018)014} {\bibfield
  {journal} {\bibinfo  {journal} {J. High Energy Phys.}\ }\textbf {\bibinfo
  {volume} {07}},\ \bibinfo {pages} {014}}\BibitemShut {NoStop}%
\bibitem [{\citenamefont {Billam}\ \emph {et~al.}(2020)\citenamefont {Billam},
  \citenamefont {Brown},\ and\ \citenamefont {Moss}}]{Billam_20}%
  \BibitemOpen
  \bibfield  {author} {\bibinfo {author} {\bibfnamefont {T.~P.}\ \bibnamefont
  {Billam}}, \bibinfo {author} {\bibfnamefont {K.}~\bibnamefont {Brown}},\ and\
  \bibinfo {author} {\bibfnamefont {I.~G.}\ \bibnamefont {Moss}},\ }\bibfield
  {title} {\bibinfo {title} {Simulating cosmological supercooling with a
  cold-atom system},\ }\href {https://doi.org/10.1103/PhysRevA.102.043324}
  {\bibfield  {journal} {\bibinfo  {journal} {Phys. Rev. A}\ }\textbf {\bibinfo
  {volume} {102}},\ \bibinfo {pages} {043324} (\bibinfo {year}
  {2020})}\BibitemShut {NoStop}%
\bibitem [{\citenamefont {Billam}\ \emph {et~al.}(2021)\citenamefont {Billam},
  \citenamefont {Brown}, \citenamefont {Groszek},\ and\ \citenamefont
  {Moss}}]{Billam_21}%
  \BibitemOpen
  \bibfield  {author} {\bibinfo {author} {\bibfnamefont {T.~P.}\ \bibnamefont
  {Billam}}, \bibinfo {author} {\bibfnamefont {K.}~\bibnamefont {Brown}},
  \bibinfo {author} {\bibfnamefont {A.~J.}\ \bibnamefont {Groszek}},\ and\
  \bibinfo {author} {\bibfnamefont {I.~G.}\ \bibnamefont {Moss}},\ }\bibfield
  {title} {\bibinfo {title} {Simulating cosmological supercooling with a cold
  atom system. ii. thermal damping and parametric instability},\ }\href
  {https://doi.org/10.1103/PhysRevA.104.053309} {\bibfield  {journal} {\bibinfo
   {journal} {Phys. Rev. A}\ }\textbf {\bibinfo {volume} {104}},\ \bibinfo
  {pages} {053309} (\bibinfo {year} {2021})}\BibitemShut {NoStop}%
\bibitem [{\citenamefont {Billam}\ \emph {et~al.}(2022)\citenamefont {Billam},
  \citenamefont {Brown},\ and\ \citenamefont {Moss}}]{Billam_22}%
  \BibitemOpen
  \bibfield  {author} {\bibinfo {author} {\bibfnamefont {T.~P.}\ \bibnamefont
  {Billam}}, \bibinfo {author} {\bibfnamefont {K.}~\bibnamefont {Brown}},\ and\
  \bibinfo {author} {\bibfnamefont {I.~G.}\ \bibnamefont {Moss}},\ }\bibfield
  {title} {\bibinfo {title} {False-vacuum decay in an ultracold spin-1 {B}ose
  gas},\ }\href {https://doi.org/10.1103/PhysRevA.105.L041301} {\bibfield
  {journal} {\bibinfo  {journal} {Phys. Rev. A}\ }\textbf {\bibinfo {volume}
  {105}},\ \bibinfo {pages} {L041301} (\bibinfo {year} {2022})}\BibitemShut
  {NoStop}%
\bibitem [{\citenamefont {Zenesini}\ \emph {et~al.}(2024)\citenamefont
  {Zenesini}, \citenamefont {Berti}, \citenamefont {Cominotti}, \citenamefont
  {Rogora}, \citenamefont {Moss}, \citenamefont {Billam}, \citenamefont
  {Carusotto}, \citenamefont {Lamporesi}, \citenamefont {Recati},\ and\
  \citenamefont {Ferrari}}]{zenesini_24}%
  \BibitemOpen
  \bibfield  {author} {\bibinfo {author} {\bibfnamefont {A.}~\bibnamefont
  {Zenesini}}, \bibinfo {author} {\bibfnamefont {A.}~\bibnamefont {Berti}},
  \bibinfo {author} {\bibfnamefont {R.}~\bibnamefont {Cominotti}}, \bibinfo
  {author} {\bibfnamefont {C.}~\bibnamefont {Rogora}}, \bibinfo {author}
  {\bibfnamefont {I.~G.}\ \bibnamefont {Moss}}, \bibinfo {author}
  {\bibfnamefont {T.~P.}\ \bibnamefont {Billam}}, \bibinfo {author}
  {\bibfnamefont {I.}~\bibnamefont {Carusotto}}, \bibinfo {author}
  {\bibfnamefont {G.}~\bibnamefont {Lamporesi}}, \bibinfo {author}
  {\bibfnamefont {A.}~\bibnamefont {Recati}},\ and\ \bibinfo {author}
  {\bibfnamefont {G.}~\bibnamefont {Ferrari}},\ }\bibfield  {title} {\bibinfo
  {title} {False vacuum decay via bubble formation in ferromagnetic
  superfluids},\ }\href {https://doi.org/10.1038/s41567-023-02345-4} {\bibfield
   {journal} {\bibinfo  {journal} {Nat. Phys.}\ }\textbf {\bibinfo {volume}
  {20}},\ \bibinfo {pages} {558} (\bibinfo {year} {2024})}\BibitemShut
  {NoStop}%
\bibitem [{\citenamefont {Darbha}\ \emph {et~al.}(2024)\citenamefont {Darbha},
  \citenamefont {Kornja\v{c}a}, \citenamefont {Liu}, \citenamefont {Balewski},
  \citenamefont {Hirsbrunner}, \citenamefont {Lopes}, \citenamefont {Wang},
  \citenamefont {Van~Beeumen}, \citenamefont {Camps},\ and\ \citenamefont
  {Klymko}}]{Darbha_24}%
  \BibitemOpen
  \bibfield  {author} {\bibinfo {author} {\bibfnamefont {S.}~\bibnamefont
  {Darbha}}, \bibinfo {author} {\bibfnamefont {M.}~\bibnamefont
  {Kornja\v{c}a}}, \bibinfo {author} {\bibfnamefont {F.}~\bibnamefont {Liu}},
  \bibinfo {author} {\bibfnamefont {J.}~\bibnamefont {Balewski}}, \bibinfo
  {author} {\bibfnamefont {M.~R.}\ \bibnamefont {Hirsbrunner}}, \bibinfo
  {author} {\bibfnamefont {P.~L.~S.}\ \bibnamefont {Lopes}}, \bibinfo {author}
  {\bibfnamefont {S.-T.}\ \bibnamefont {Wang}}, \bibinfo {author}
  {\bibfnamefont {R.}~\bibnamefont {Van~Beeumen}}, \bibinfo {author}
  {\bibfnamefont {D.}~\bibnamefont {Camps}},\ and\ \bibinfo {author}
  {\bibfnamefont {K.}~\bibnamefont {Klymko}},\ }\bibfield  {title} {\bibinfo
  {title} {False vacuum decay and nucleation dynamics in neutral atom
  systems},\ }\href {https://doi.org/10.1103/PhysRevB.110.155103} {\bibfield
  {journal} {\bibinfo  {journal} {Phys. Rev. B}\ }\textbf {\bibinfo {volume}
  {110}},\ \bibinfo {pages} {155103} (\bibinfo {year} {2024})}\BibitemShut
  {NoStop}%
\bibitem [{\citenamefont {Brown}\ \emph {et~al.}(2025)\citenamefont {Brown},
  \citenamefont {Moss},\ and\ \citenamefont {Billam}}]{K.brown_25}%
  \BibitemOpen
  \bibfield  {author} {\bibinfo {author} {\bibfnamefont {K.}~\bibnamefont
  {Brown}}, \bibinfo {author} {\bibfnamefont {I.~G.}\ \bibnamefont {Moss}},\
  and\ \bibinfo {author} {\bibfnamefont {T.~P.}\ \bibnamefont {Billam}},\
  }\href@noop {} {\bibinfo {title} {Mitigating boundary effects in finite
  temperature simulations of false vacuum decay}} (\bibinfo {year} {2025}),\
  \Eprint {https://arxiv.org/abs/2504.03509} {arXiv:2504.03509} \BibitemShut
  {NoStop}%
\bibitem [{\citenamefont {Jenkins}\ \emph {et~al.}(2025)\citenamefont
  {Jenkins}, \citenamefont {Peiris},\ and\ \citenamefont
  {Pontzen}}]{jenkins_25}%
  \BibitemOpen
  \bibfield  {author} {\bibinfo {author} {\bibfnamefont {A.~C.}\ \bibnamefont
  {Jenkins}}, \bibinfo {author} {\bibfnamefont {H.~V.}\ \bibnamefont
  {Peiris}},\ and\ \bibinfo {author} {\bibfnamefont {A.}~\bibnamefont
  {Pontzen}},\ }\bibfield  {title} {\bibinfo {title} {{B}ubbles in a box:
  {E}liminating edge nucleation in cold-atom simulators of vacuum decay},\
  }\href {https://doi.org/10.1103/lrc6-49q1} {\bibfield  {journal} {\bibinfo
  {journal} {Phys. Rev. A}\ }\textbf {\bibinfo {volume} {112}},\ \bibinfo
  {pages} {023318} (\bibinfo {year} {2025})}\BibitemShut {NoStop}%
\bibitem [{\citenamefont {Recati}\ and\ \citenamefont
  {Stringari}(2022)}]{Alessio_22}%
  \BibitemOpen
  \bibfield  {author} {\bibinfo {author} {\bibfnamefont {A.}~\bibnamefont
  {Recati}}\ and\ \bibinfo {author} {\bibfnamefont {S.}~\bibnamefont
  {Stringari}},\ }\bibfield  {title} {\bibinfo {title} {Coherently coupled
  mixtures of ultracold atomic gases},\ }\href
  {https://doi.org/10.1146/annurev-conmatphys-031820-121316} {\bibfield
  {journal} {\bibinfo  {journal} {Annu. Rev. Condens. Matter Phys.}\ }\textbf
  {\bibinfo {volume} {13}},\ \bibinfo {pages} {407} (\bibinfo {year}
  {2022})}\BibitemShut {NoStop}%
\bibitem [{\citenamefont {Cominotti}\ \emph {et~al.}(2025)\citenamefont
  {Cominotti}, \citenamefont {Baroni}, \citenamefont {Rogora}, \citenamefont
  {Andreoni}, \citenamefont {Guarda}, \citenamefont {Lamporesi}, \citenamefont
  {Ferrari},\ and\ \citenamefont {Zenesini}}]{cominotti_25}%
  \BibitemOpen
  \bibfield  {author} {\bibinfo {author} {\bibfnamefont {R.}~\bibnamefont
  {Cominotti}}, \bibinfo {author} {\bibfnamefont {C.}~\bibnamefont {Baroni}},
  \bibinfo {author} {\bibfnamefont {C.}~\bibnamefont {Rogora}}, \bibinfo
  {author} {\bibfnamefont {D.}~\bibnamefont {Andreoni}}, \bibinfo {author}
  {\bibfnamefont {G.}~\bibnamefont {Guarda}}, \bibinfo {author} {\bibfnamefont
  {G.}~\bibnamefont {Lamporesi}}, \bibinfo {author} {\bibfnamefont
  {G.}~\bibnamefont {Ferrari}},\ and\ \bibinfo {author} {\bibfnamefont
  {A.}~\bibnamefont {Zenesini}},\ }\bibfield  {title} {\bibinfo {title}
  {Observation of temperature effects on false vacuum decay in atomic quantum
  gases},\ }\href {https://doi.org/10.1103/l396-yysb} {\bibfield  {journal}
  {\bibinfo  {journal} {Phys. Rev. Lett.}\ }\textbf {\bibinfo {volume} {135}},\
  \bibinfo {pages} {183401} (\bibinfo {year} {2025})}\BibitemShut {NoStop}%
\bibitem [{\citenamefont {Zibold}\ \emph {et~al.}(2010)\citenamefont {Zibold},
  \citenamefont {Nicklas}, \citenamefont {Gross},\ and\ \citenamefont
  {Oberthaler}}]{T.Zibold_10}%
  \BibitemOpen
  \bibfield  {author} {\bibinfo {author} {\bibfnamefont {T.}~\bibnamefont
  {Zibold}}, \bibinfo {author} {\bibfnamefont {E.}~\bibnamefont {Nicklas}},
  \bibinfo {author} {\bibfnamefont {C.}~\bibnamefont {Gross}},\ and\ \bibinfo
  {author} {\bibfnamefont {M.~K.}\ \bibnamefont {Oberthaler}},\ }\bibfield
  {title} {\bibinfo {title} {Classical bifurcation at the transition from
  {R}abi to {J}osephson dynamics},\ }\href
  {https://doi.org/10.1103/PhysRevLett.105.204101} {\bibfield  {journal}
  {\bibinfo  {journal} {Phys. Rev. Lett.}\ }\textbf {\bibinfo {volume} {105}},\
  \bibinfo {pages} {204101} (\bibinfo {year} {2010})}\BibitemShut {NoStop}%
\bibitem [{\citenamefont {Farolfi}\ \emph
  {et~al.}(2021{\natexlab{a}})\citenamefont {Farolfi}, \citenamefont
  {Zenesini}, \citenamefont {Trypogeorgos}, \citenamefont {Mordini},
  \citenamefont {Gallem{\'i}}, \citenamefont {Roy}, \citenamefont {Recati},
  \citenamefont {Lamporesi},\ and\ \citenamefont
  {Ferrari}}]{farolfi_2021_torque}%
  \BibitemOpen
  \bibfield  {author} {\bibinfo {author} {\bibfnamefont {A.}~\bibnamefont
  {Farolfi}}, \bibinfo {author} {\bibfnamefont {A.}~\bibnamefont {Zenesini}},
  \bibinfo {author} {\bibfnamefont {D.}~\bibnamefont {Trypogeorgos}}, \bibinfo
  {author} {\bibfnamefont {C.}~\bibnamefont {Mordini}}, \bibinfo {author}
  {\bibfnamefont {A.}~\bibnamefont {Gallem{\'i}}}, \bibinfo {author}
  {\bibfnamefont {A.}~\bibnamefont {Roy}}, \bibinfo {author} {\bibfnamefont
  {A.}~\bibnamefont {Recati}}, \bibinfo {author} {\bibfnamefont
  {G.}~\bibnamefont {Lamporesi}},\ and\ \bibinfo {author} {\bibfnamefont
  {G.}~\bibnamefont {Ferrari}},\ }\bibfield  {title} {\bibinfo {title}
  {Quantum-torque-induced breaking of magnetic interfaces in ultracold gases},\
  }\href {https://doi.org/10.1038/s41567-021-01369-y} {\bibfield  {journal}
  {\bibinfo  {journal} {Nat. Phys.}\ }\textbf {\bibinfo {volume} {17}},\
  \bibinfo {pages} {1359} (\bibinfo {year} {2021}{\natexlab{a}})}\BibitemShut
  {NoStop}%
\bibitem [{\citenamefont {Farolfi}\ \emph
  {et~al.}(2021{\natexlab{b}})\citenamefont {Farolfi}, \citenamefont
  {Zenesini}, \citenamefont {Cominotti}, \citenamefont {Trypogeorgos},
  \citenamefont {Recati}, \citenamefont {Lamporesi},\ and\ \citenamefont
  {Ferrari}}]{Farolfi_21}%
  \BibitemOpen
  \bibfield  {author} {\bibinfo {author} {\bibfnamefont {A.}~\bibnamefont
  {Farolfi}}, \bibinfo {author} {\bibfnamefont {A.}~\bibnamefont {Zenesini}},
  \bibinfo {author} {\bibfnamefont {R.}~\bibnamefont {Cominotti}}, \bibinfo
  {author} {\bibfnamefont {D.}~\bibnamefont {Trypogeorgos}}, \bibinfo {author}
  {\bibfnamefont {A.}~\bibnamefont {Recati}}, \bibinfo {author} {\bibfnamefont
  {G.}~\bibnamefont {Lamporesi}},\ and\ \bibinfo {author} {\bibfnamefont
  {G.}~\bibnamefont {Ferrari}},\ }\bibfield  {title} {\bibinfo {title}
  {Manipulation of an elongated internal {J}osephson junction of bosonic
  atoms},\ }\href {https://doi.org/10.1103/PhysRevA.104.023326} {\bibfield
  {journal} {\bibinfo  {journal} {Phys. Rev. A}\ }\textbf {\bibinfo {volume}
  {104}},\ \bibinfo {pages} {023326} (\bibinfo {year}
  {2021}{\natexlab{b}})}\BibitemShut {NoStop}%
\bibitem [{\citenamefont {Eto}\ and\ \citenamefont
  {Nitta}(2018)}]{MinoruEto_18}%
  \BibitemOpen
  \bibfield  {author} {\bibinfo {author} {\bibfnamefont {M.}~\bibnamefont
  {Eto}}\ and\ \bibinfo {author} {\bibfnamefont {M.}~\bibnamefont {Nitta}},\
  }\bibfield  {title} {\bibinfo {title} {Confinement of half-quantized vortices
  in coherently coupled {B}ose-{E}instein condensates: Simulating quark
  confinement in a qcd-like theory},\ }\href
  {https://doi.org/10.1103/PhysRevA.97.023613} {\bibfield  {journal} {\bibinfo
  {journal} {Phys. Rev. A}\ }\textbf {\bibinfo {volume} {97}},\ \bibinfo
  {pages} {023613} (\bibinfo {year} {2018})}\BibitemShut {NoStop}%
\bibitem [{\citenamefont {Cominotti}\ \emph {et~al.}(2023)\citenamefont
  {Cominotti}, \citenamefont {Berti}, \citenamefont {Dulin}, \citenamefont
  {Rogora}, \citenamefont {Lamporesi}, \citenamefont {Carusotto}, \citenamefont
  {Recati}, \citenamefont {Zenesini},\ and\ \citenamefont
  {Ferrari}}]{Cominotti_23}%
  \BibitemOpen
  \bibfield  {author} {\bibinfo {author} {\bibfnamefont {R.}~\bibnamefont
  {Cominotti}}, \bibinfo {author} {\bibfnamefont {A.}~\bibnamefont {Berti}},
  \bibinfo {author} {\bibfnamefont {C.}~\bibnamefont {Dulin}}, \bibinfo
  {author} {\bibfnamefont {C.}~\bibnamefont {Rogora}}, \bibinfo {author}
  {\bibfnamefont {G.}~\bibnamefont {Lamporesi}}, \bibinfo {author}
  {\bibfnamefont {I.}~\bibnamefont {Carusotto}}, \bibinfo {author}
  {\bibfnamefont {A.}~\bibnamefont {Recati}}, \bibinfo {author} {\bibfnamefont
  {A.}~\bibnamefont {Zenesini}},\ and\ \bibinfo {author} {\bibfnamefont
  {G.}~\bibnamefont {Ferrari}},\ }\bibfield  {title} {\bibinfo {title}
  {Ferromagnetism in an extended coherently coupled atomic superfluid},\ }\href
  {https://doi.org/10.1103/PhysRevX.13.021037} {\bibfield  {journal} {\bibinfo
  {journal} {Phys. Rev. X}\ }\textbf {\bibinfo {volume} {13}},\ \bibinfo
  {pages} {021037} (\bibinfo {year} {2023})}\BibitemShut {NoStop}%
\bibitem [{\citenamefont {Linde}(1983)}]{Linde_83}%
  \BibitemOpen
  \bibfield  {author} {\bibinfo {author} {\bibfnamefont {A.~D.}\ \bibnamefont
  {Linde}},\ }\bibfield  {title} {\bibinfo {title} {Decay of the false vacuum
  at finite temperature},\ }\href
  {https://doi.org/10.1016/0550-3213(83)90072-X} {\bibfield  {journal}
  {\bibinfo  {journal} {Nucl. Phys. B}\ }\textbf {\bibinfo {volume} {216}},\
  \bibinfo {pages} {421} (\bibinfo {year} {1983})},\ \bibinfo {note} {[Erratum:
  Nucl. Phys. B 223, 544 (1983)]}\BibitemShut {NoStop}%
\bibitem [{\citenamefont {Stoof}\ and\ \citenamefont
  {Bijlsma}(2001)}]{Stoof_01}%
  \BibitemOpen
  \bibfield  {author} {\bibinfo {author} {\bibfnamefont {H.~T.~C.}\
  \bibnamefont {Stoof}}\ and\ \bibinfo {author} {\bibfnamefont {M.~J.}\
  \bibnamefont {Bijlsma}},\ }\bibfield  {title} {\bibinfo {title} {Dynamics of
  fluctuating {B}ose–{E}instein condensates},\ }\href
  {https://doi.org/10.1023/A:1017519118408} {\bibfield  {journal} {\bibinfo
  {journal} {J. Low Temp. Phys.}\ }\textbf {\bibinfo {volume} {124}},\ \bibinfo
  {pages} {431} (\bibinfo {year} {2001})}\BibitemShut {NoStop}%
\bibitem [{\citenamefont {Proukakis}\ and\ \citenamefont
  {Jackson}(2008)}]{Proukakis_08}%
  \BibitemOpen
  \bibfield  {author} {\bibinfo {author} {\bibfnamefont {N.~P.}\ \bibnamefont
  {Proukakis}}\ and\ \bibinfo {author} {\bibfnamefont {B.}~\bibnamefont
  {Jackson}},\ }\bibfield  {title} {\bibinfo {title} {Finite-temperature models
  of {B}ose-{E}instein condensation},\ }\href
  {https://doi.org/10.1088/0953-4075/41/20/203002} {\bibfield  {journal}
  {\bibinfo  {journal} {J. Phys. B: At. Mol. Opt. Phys.}\ }\textbf {\bibinfo
  {volume} {41}},\ \bibinfo {pages} {203002} (\bibinfo {year}
  {2008})}\BibitemShut {NoStop}%
\bibitem [{\citenamefont {Blakie}\ \emph {et~al.}(2008)\citenamefont {Blakie},
  \citenamefont {Bradley}, \citenamefont {Davis}, \citenamefont {Ballagh},\
  and\ \citenamefont {Gardiner}}]{Blakie_08}%
  \BibitemOpen
  \bibfield  {author} {\bibinfo {author} {\bibfnamefont {P.~B.}\ \bibnamefont
  {Blakie}}, \bibinfo {author} {\bibfnamefont {A.~S.}\ \bibnamefont {Bradley}},
  \bibinfo {author} {\bibfnamefont {M.~J.}\ \bibnamefont {Davis}}, \bibinfo
  {author} {\bibfnamefont {R.~J.}\ \bibnamefont {Ballagh}},\ and\ \bibinfo
  {author} {\bibfnamefont {C.~W.}\ \bibnamefont {Gardiner}},\ }\bibfield
  {title} {\bibinfo {title} {Dynamics and statistical mechanics of ultra-cold
  {B}ose gases using c-field techniques},\ }\href
  {https://doi.org/10.1080/00018730802564254} {\bibfield  {journal} {\bibinfo
  {journal} {Adv. Phys.}\ }\textbf {\bibinfo {volume} {57}},\ \bibinfo {pages}
  {363} (\bibinfo {year} {2008})}\BibitemShut {NoStop}%
\bibitem [{\citenamefont {Baroni}\ \emph {et~al.}(2024)\citenamefont {Baroni},
  \citenamefont {Lamporesi},\ and\ \citenamefont {Zaccanti}}]{baroni_24}%
  \BibitemOpen
  \bibfield  {author} {\bibinfo {author} {\bibfnamefont {C.}~\bibnamefont
  {Baroni}}, \bibinfo {author} {\bibfnamefont {G.}~\bibnamefont {Lamporesi}},\
  and\ \bibinfo {author} {\bibfnamefont {M.}~\bibnamefont {Zaccanti}},\
  }\bibfield  {title} {\bibinfo {title} {Quantum mixtures of ultracold gases of
  neutral atoms},\ }\href {https://doi.org/10.1038/s42254-024-00773-6}
  {\bibfield  {journal} {\bibinfo  {journal} {Nat. Rev. Phys.}\ }\textbf
  {\bibinfo {volume} {6}},\ \bibinfo {pages} {736} (\bibinfo {year}
  {2024})}\BibitemShut {NoStop}%
\bibitem [{\citenamefont {Farolfi}(2021)}]{farolfi_phd21}%
  \BibitemOpen
  \bibfield  {author} {\bibinfo {author} {\bibfnamefont {A.}~\bibnamefont
  {Farolfi}},\ }\href {https://iris.unitn.it/handle/11572/299835} {\bibinfo
  {title} {Spin dynamics in two-component {B}ose-{E}instein condensates.
  {T}hesis, {U}niversity of {T}rento.}} (\bibinfo {year} {2021})\BibitemShut
  {NoStop}%
\bibitem [{\citenamefont {Roy}\ \emph {et~al.}(2023)\citenamefont {Roy},
  \citenamefont {Ota}, \citenamefont {Dalfovo},\ and\ \citenamefont
  {Recati}}]{Arko_23}%
  \BibitemOpen
  \bibfield  {author} {\bibinfo {author} {\bibfnamefont {A.}~\bibnamefont
  {Roy}}, \bibinfo {author} {\bibfnamefont {M.}~\bibnamefont {Ota}}, \bibinfo
  {author} {\bibfnamefont {F.}~\bibnamefont {Dalfovo}},\ and\ \bibinfo {author}
  {\bibfnamefont {A.}~\bibnamefont {Recati}},\ }\bibfield  {title} {\bibinfo
  {title} {Finite-temperature ferromagnetic transition in coherently coupled
  {B}ose gases},\ }\href {https://doi.org/10.1103/PhysRevA.107.043301}
  {\bibfield  {journal} {\bibinfo  {journal} {Phys. Rev. A}\ }\textbf {\bibinfo
  {volume} {107}},\ \bibinfo {pages} {043301} (\bibinfo {year}
  {2023})}\BibitemShut {NoStop}%
\bibitem [{\citenamefont {Bradley}\ and\ \citenamefont
  {Blakie}(2014)}]{A.bradley_14}%
  \BibitemOpen
  \bibfield  {author} {\bibinfo {author} {\bibfnamefont {A.~S.}\ \bibnamefont
  {Bradley}}\ and\ \bibinfo {author} {\bibfnamefont {P.~B.}\ \bibnamefont
  {Blakie}},\ }\bibfield  {title} {\bibinfo {title} {Stochastic projected
  {G}ross-{P}itaevskii equation for spinor and multicomponent condensates},\
  }\href {https://doi.org/10.1103/PhysRevA.90.023631} {\bibfield  {journal}
  {\bibinfo  {journal} {Phys. Rev. A}\ }\textbf {\bibinfo {volume} {90}},\
  \bibinfo {pages} {023631} (\bibinfo {year} {2014})}\BibitemShut {NoStop}%
\bibitem [{\citenamefont {Liu}\ \emph {et~al.}(2016)\citenamefont {Liu},
  \citenamefont {Pattinson}, \citenamefont {Billam}, \citenamefont {Gardiner},
  \citenamefont {Cornish}, \citenamefont {Huang}, \citenamefont {Lin},
  \citenamefont {Gou}, \citenamefont {Parker},\ and\ \citenamefont
  {Proukakis}}]{Liu_16}%
  \BibitemOpen
  \bibfield  {author} {\bibinfo {author} {\bibfnamefont {I.-K.}\ \bibnamefont
  {Liu}}, \bibinfo {author} {\bibfnamefont {R.~W.}\ \bibnamefont {Pattinson}},
  \bibinfo {author} {\bibfnamefont {T.~P.}\ \bibnamefont {Billam}}, \bibinfo
  {author} {\bibfnamefont {S.~A.}\ \bibnamefont {Gardiner}}, \bibinfo {author}
  {\bibfnamefont {S.~L.}\ \bibnamefont {Cornish}}, \bibinfo {author}
  {\bibfnamefont {T.-M.}\ \bibnamefont {Huang}}, \bibinfo {author}
  {\bibfnamefont {W.-W.}\ \bibnamefont {Lin}}, \bibinfo {author} {\bibfnamefont
  {S.-C.}\ \bibnamefont {Gou}}, \bibinfo {author} {\bibfnamefont {N.~G.}\
  \bibnamefont {Parker}},\ and\ \bibinfo {author} {\bibfnamefont {N.~P.}\
  \bibnamefont {Proukakis}},\ }\bibfield  {title} {\bibinfo {title}
  {{S}tochastic growth dynamics and composite defects in quenched immiscible
  binary condensates},\ }\href {https://doi.org/10.1103/PhysRevA.93.023628}
  {\bibfield  {journal} {\bibinfo  {journal} {Phys. Rev. A}\ }\textbf {\bibinfo
  {volume} {93}},\ \bibinfo {pages} {023628} (\bibinfo {year}
  {2016})}\BibitemShut {NoStop}%
\bibitem [{\citenamefont {Ota}\ \emph {et~al.}(2018)\citenamefont {Ota},
  \citenamefont {Larcher}, \citenamefont {Dalfovo}, \citenamefont {Pitaevskii},
  \citenamefont {Proukakis},\ and\ \citenamefont {Stringari}}]{miki_ota_18}%
  \BibitemOpen
  \bibfield  {author} {\bibinfo {author} {\bibfnamefont {M.}~\bibnamefont
  {Ota}}, \bibinfo {author} {\bibfnamefont {F.}~\bibnamefont {Larcher}},
  \bibinfo {author} {\bibfnamefont {F.}~\bibnamefont {Dalfovo}}, \bibinfo
  {author} {\bibfnamefont {L.}~\bibnamefont {Pitaevskii}}, \bibinfo {author}
  {\bibfnamefont {N.~P.}\ \bibnamefont {Proukakis}},\ and\ \bibinfo {author}
  {\bibfnamefont {S.}~\bibnamefont {Stringari}},\ }\bibfield  {title} {\bibinfo
  {title} {Collisionless sound in a uniform two-dimensional {B}ose gas},\
  }\href {https://doi.org/10.1103/PhysRevLett.121.145302} {\bibfield  {journal}
  {\bibinfo  {journal} {Phys. Rev. Lett.}\ }\textbf {\bibinfo {volume} {121}},\
  \bibinfo {pages} {145302} (\bibinfo {year} {2018})}\BibitemShut {NoStop}%
\bibitem [{\citenamefont {Roy}\ \emph {et~al.}(2021)\citenamefont {Roy},
  \citenamefont {Ota}, \citenamefont {Recati},\ and\ \citenamefont
  {Dalfovo}}]{Arko_21}%
  \BibitemOpen
  \bibfield  {author} {\bibinfo {author} {\bibfnamefont {A.}~\bibnamefont
  {Roy}}, \bibinfo {author} {\bibfnamefont {M.}~\bibnamefont {Ota}}, \bibinfo
  {author} {\bibfnamefont {A.}~\bibnamefont {Recati}},\ and\ \bibinfo {author}
  {\bibfnamefont {F.}~\bibnamefont {Dalfovo}},\ }\bibfield  {title} {\bibinfo
  {title} {Finite-temperature spin dynamics of a two-dimensional {B}ose-{B}ose
  atomic mixture},\ }\href {https://doi.org/10.1103/PhysRevResearch.3.013161}
  {\bibfield  {journal} {\bibinfo  {journal} {Phys. Rev. Res.}\ }\textbf
  {\bibinfo {volume} {3}},\ \bibinfo {pages} {013161} (\bibinfo {year}
  {2021})}\BibitemShut {NoStop}%
\bibitem [{\citenamefont {Larcher}(2018)}]{Larcher2018}%
  \BibitemOpen
  \bibfield  {author} {\bibinfo {author} {\bibfnamefont {F.}~\bibnamefont
  {Larcher}},\ }\href {https://iris.unitn.it/handle/11572/368598} {\bibinfo
  {title} {Dynamical excitations in low-dimensional condensates: Sound,
  vortices, and quenched dynamics. {T}hesis, {U}niversity of {T}rento and
  {N}ewcastle {U}niversity.}} (\bibinfo {year} {2018})\BibitemShut {NoStop}%
\bibitem [{\citenamefont {Billam}\ \emph {et~al.}(2023)\citenamefont {Billam},
  \citenamefont {Brown},\ and\ \citenamefont {Moss}}]{Billam_2023}%
  \BibitemOpen
  \bibfield  {author} {\bibinfo {author} {\bibfnamefont {T.~P.}\ \bibnamefont
  {Billam}}, \bibinfo {author} {\bibfnamefont {K.}~\bibnamefont {Brown}},\ and\
  \bibinfo {author} {\bibfnamefont {I.~G.}\ \bibnamefont {Moss}},\ }\bibfield
  {title} {\bibinfo {title} {Bubble nucleation in a cold spin 1 gas},\ }\href
  {https://doi.org/10.1088/1367-2630/accca2} {\bibfield  {journal} {\bibinfo
  {journal} {New Journal of Physics}\ }\textbf {\bibinfo {volume} {25}},\
  \bibinfo {pages} {043028} (\bibinfo {year} {2023})}\BibitemShut {NoStop}%
\bibitem [{\citenamefont {Liu}\ \emph {et~al.}(2012)\citenamefont {Liu},
  \citenamefont {Fan}, \citenamefont {Zhang}, \citenamefont {Wang},\ and\
  \citenamefont {Liu}}]{liu_12}%
  \BibitemOpen
  \bibfield  {author} {\bibinfo {author} {\bibfnamefont {C.-F.}\ \bibnamefont
  {Liu}}, \bibinfo {author} {\bibfnamefont {H.}~\bibnamefont {Fan}}, \bibinfo
  {author} {\bibfnamefont {Y.-C.}\ \bibnamefont {Zhang}}, \bibinfo {author}
  {\bibfnamefont {D.-S.}\ \bibnamefont {Wang}},\ and\ \bibinfo {author}
  {\bibfnamefont {W.-M.}\ \bibnamefont {Liu}},\ }\bibfield  {title} {\bibinfo
  {title} {Circular-hyperbolic skyrmion in rotating pseudo-spin-1/2
  {B}ose-{E}instein condensates with spin-orbit coupling},\ }\href
  {https://doi.org/10.1103/PhysRevA.86.053616} {\bibfield  {journal} {\bibinfo
  {journal} {Phys. Rev. A}\ }\textbf {\bibinfo {volume} {86}},\ \bibinfo
  {pages} {053616} (\bibinfo {year} {2012})}\BibitemShut {NoStop}%
\bibitem [{\citenamefont {Su}\ \emph {et~al.}(2012)\citenamefont {Su},
  \citenamefont {Liu}, \citenamefont {Tsai}, \citenamefont {Liu},\ and\
  \citenamefont {Gou}}]{su_12}%
  \BibitemOpen
  \bibfield  {author} {\bibinfo {author} {\bibfnamefont {S.-W.}\ \bibnamefont
  {Su}}, \bibinfo {author} {\bibfnamefont {I.-K.}\ \bibnamefont {Liu}},
  \bibinfo {author} {\bibfnamefont {Y.-C.}\ \bibnamefont {Tsai}}, \bibinfo
  {author} {\bibfnamefont {W.~M.}\ \bibnamefont {Liu}},\ and\ \bibinfo {author}
  {\bibfnamefont {S.-C.}\ \bibnamefont {Gou}},\ }\bibfield  {title} {\bibinfo
  {title} {Crystallized half-skyrmions and inverted half-skyrmions in the
  condensation of spin-1 {B}ose gases with spin-orbit coupling},\ }\href
  {https://doi.org/10.1103/PhysRevA.86.023601} {\bibfield  {journal} {\bibinfo
  {journal} {Phys. Rev. A}\ }\textbf {\bibinfo {volume} {86}},\ \bibinfo
  {pages} {023601} (\bibinfo {year} {2012})}\BibitemShut {NoStop}%
\bibitem [{\citenamefont {Su}\ \emph {et~al.}(2017)\citenamefont {Su},
  \citenamefont {Liu}, \citenamefont {Gou}, \citenamefont {Liao}, \citenamefont
  {Fialko},\ and\ \citenamefont {{B}rand}}]{su_17}%
  \BibitemOpen
  \bibfield  {author} {\bibinfo {author} {\bibfnamefont {S.-W.}\ \bibnamefont
  {Su}}, \bibinfo {author} {\bibfnamefont {I.-K.}\ \bibnamefont {Liu}},
  \bibinfo {author} {\bibfnamefont {S.-C.}\ \bibnamefont {Gou}}, \bibinfo
  {author} {\bibfnamefont {R.}~\bibnamefont {Liao}}, \bibinfo {author}
  {\bibfnamefont {O.}~\bibnamefont {Fialko}},\ and\ \bibinfo {author}
  {\bibfnamefont {J.}~\bibnamefont {{B}rand}},\ }\bibfield  {title} {\bibinfo
  {title} {Hidden long-range order in a spin-orbit-coupled two-dimensional
  {B}ose gas},\ }\href {https://doi.org/10.1103/PhysRevA.95.053629} {\bibfield
  {journal} {\bibinfo  {journal} {Phys. Rev. A}\ }\textbf {\bibinfo {volume}
  {95}},\ \bibinfo {pages} {053629} (\bibinfo {year} {2017})}\BibitemShut
  {NoStop}%
\bibitem [{\citenamefont {Sakmann}\ and\ \citenamefont
  {Kasevich}(2016)}]{Sakmann_16}%
  \BibitemOpen
  \bibfield  {author} {\bibinfo {author} {\bibfnamefont {K.}~\bibnamefont
  {Sakmann}}\ and\ \bibinfo {author} {\bibfnamefont {M.}~\bibnamefont
  {Kasevich}},\ }\bibfield  {title} {\bibinfo {title} {{S}ingle-shot
  simulations of dynamic quantum many-body systems},\ }\href
  {https://doi.org/10.1038/nphys3631} {\bibfield  {journal} {\bibinfo
  {journal} {Nat. Phys.}\ }\textbf {\bibinfo {volume} {12}},\ \bibinfo {pages}
  {451} (\bibinfo {year} {2016})}\BibitemShut {NoStop}%
\bibitem [{\citenamefont {Sakmann}\ and\ \citenamefont
  {Kasevich}(2017)}]{Sakmann_17}%
  \BibitemOpen
  \bibfield  {author} {\bibinfo {author} {\bibfnamefont {K.}~\bibnamefont
  {Sakmann}}\ and\ \bibinfo {author} {\bibfnamefont {M.}~\bibnamefont
  {Kasevich}},\ }\href@noop {} {\bibinfo {title} {Reply to the correspondence
  of {D}rummond and {B}rand [arxiv:1610.07633]}} (\bibinfo {year} {2017}),\
  \Eprint {https://arxiv.org/abs/1702.01211} {arXiv:1702.01211} \BibitemShut
  {NoStop}%
\bibitem [{\citenamefont {Olsen}\ \emph {et~al.}(2017)\citenamefont {Olsen},
  \citenamefont {Corney}, \citenamefont {Lewis-Swan},\ and\ \citenamefont
  {Bradley}}]{Olsen_17}%
  \BibitemOpen
  \bibfield  {author} {\bibinfo {author} {\bibfnamefont {M.~K.}\ \bibnamefont
  {Olsen}}, \bibinfo {author} {\bibfnamefont {J.~F.}\ \bibnamefont {Corney}},
  \bibinfo {author} {\bibfnamefont {R.~J.}\ \bibnamefont {Lewis-Swan}},\ and\
  \bibinfo {author} {\bibfnamefont {A.~S.}\ \bibnamefont {Bradley}},\
  }\href@noop {} {\bibinfo {title} {Correspondence on ``{S}ingle-shot
  simulations of dynamic quantum many-body systems''}} (\bibinfo {year}
  {2017}),\ \Eprint {https://arxiv.org/abs/1702.00282} {arXiv:1702.00282}
  \BibitemShut {NoStop}%
\bibitem [{\citenamefont {Gaunt}\ \emph {et~al.}(2013)\citenamefont {Gaunt},
  \citenamefont {Schmidutz}, \citenamefont {Gotlibovych}, \citenamefont
  {Smith},\ and\ \citenamefont {Hadzibabic}}]{gaunt_13}%
  \BibitemOpen
  \bibfield  {author} {\bibinfo {author} {\bibfnamefont {A.~L.}\ \bibnamefont
  {Gaunt}}, \bibinfo {author} {\bibfnamefont {T.~F.}\ \bibnamefont
  {Schmidutz}}, \bibinfo {author} {\bibfnamefont {I.}~\bibnamefont
  {Gotlibovych}}, \bibinfo {author} {\bibfnamefont {R.~P.}\ \bibnamefont
  {Smith}},\ and\ \bibinfo {author} {\bibfnamefont {Z.}~\bibnamefont
  {Hadzibabic}},\ }\bibfield  {title} {\bibinfo {title} {{B}ose-{E}instein
  condensation of atoms in a uniform potential},\ }\href
  {https://doi.org/10.1103/PhysRevLett.110.200406} {\bibfield  {journal}
  {\bibinfo  {journal} {Phys. Rev. Lett.}\ }\textbf {\bibinfo {volume} {110}},\
  \bibinfo {pages} {200406} (\bibinfo {year} {2013})}\BibitemShut {NoStop}%
\bibitem [{\citenamefont {Ville}\ \emph {et~al.}(2018)\citenamefont {Ville},
  \citenamefont {Saint-Jalm}, \citenamefont {Le~Cerf}, \citenamefont
  {Aidelsburger}, \citenamefont {Nascimb\`ene}, \citenamefont {Dalibard},\ and\
  \citenamefont {Beugnon}}]{Ville_18}%
  \BibitemOpen
  \bibfield  {author} {\bibinfo {author} {\bibfnamefont {J.~L.}\ \bibnamefont
  {Ville}}, \bibinfo {author} {\bibfnamefont {R.}~\bibnamefont {Saint-Jalm}},
  \bibinfo {author} {\bibfnamefont {E.}~\bibnamefont {Le~Cerf}}, \bibinfo
  {author} {\bibfnamefont {M.}~\bibnamefont {Aidelsburger}}, \bibinfo {author}
  {\bibfnamefont {S.}~\bibnamefont {Nascimb\`ene}}, \bibinfo {author}
  {\bibfnamefont {J.}~\bibnamefont {Dalibard}},\ and\ \bibinfo {author}
  {\bibfnamefont {J.}~\bibnamefont {Beugnon}},\ }\bibfield  {title} {\bibinfo
  {title} {Sound propagation in a uniform superfluid two-dimensional {B}ose
  gas},\ }\href {https://doi.org/10.1103/PhysRevLett.121.145301} {\bibfield
  {journal} {\bibinfo  {journal} {Phys. Rev. Lett.}\ }\textbf {\bibinfo
  {volume} {121}},\ \bibinfo {pages} {145301} (\bibinfo {year}
  {2018})}\BibitemShut {NoStop}%
\bibitem [{\citenamefont {Cominotti}\ \emph {et~al.}(2022)\citenamefont
  {Cominotti}, \citenamefont {Berti}, \citenamefont {Farolfi}, \citenamefont
  {Zenesini}, \citenamefont {Lamporesi}, \citenamefont {Carusotto},
  \citenamefont {Recati},\ and\ \citenamefont {Ferrari}}]{cominotti_22}%
  \BibitemOpen
  \bibfield  {author} {\bibinfo {author} {\bibfnamefont {R.}~\bibnamefont
  {Cominotti}}, \bibinfo {author} {\bibfnamefont {A.}~\bibnamefont {Berti}},
  \bibinfo {author} {\bibfnamefont {A.}~\bibnamefont {Farolfi}}, \bibinfo
  {author} {\bibfnamefont {A.}~\bibnamefont {Zenesini}}, \bibinfo {author}
  {\bibfnamefont {G.}~\bibnamefont {Lamporesi}}, \bibinfo {author}
  {\bibfnamefont {I.}~\bibnamefont {Carusotto}}, \bibinfo {author}
  {\bibfnamefont {A.}~\bibnamefont {Recati}},\ and\ \bibinfo {author}
  {\bibfnamefont {G.}~\bibnamefont {Ferrari}},\ }\bibfield  {title} {\bibinfo
  {title} {Observation of massless and massive collective excitations with
  {F}araday patterns in a two-component superfluid},\ }\href
  {https://doi.org/10.1103/PhysRevLett.128.210401} {\bibfield  {journal}
  {\bibinfo  {journal} {Phys. Rev. Lett.}\ }\textbf {\bibinfo {volume} {128}},\
  \bibinfo {pages} {210401} (\bibinfo {year} {2022})}\BibitemShut {NoStop}%
\bibitem [{\citenamefont {Affleck}(1981)}]{Ian_Affleck_81}%
  \BibitemOpen
  \bibfield  {author} {\bibinfo {author} {\bibfnamefont {I.}~\bibnamefont
  {Affleck}},\ }\bibfield  {title} {\bibinfo {title} {Quantum-statistical
  metastability},\ }\href {https://doi.org/10.1103/PhysRevLett.46.388}
  {\bibfield  {journal} {\bibinfo  {journal} {Phys. Rev. Lett.}\ }\textbf
  {\bibinfo {volume} {46}},\ \bibinfo {pages} {388} (\bibinfo {year}
  {1981})}\BibitemShut {NoStop}%
\bibitem [{\citenamefont {Garcia}\ and\ \citenamefont
  {Hofmann}(2026)}]{Garcia_26}%
  \BibitemOpen
  \bibfield  {author} {\bibinfo {author} {\bibfnamefont {E.~R.}\ \bibnamefont
  {Garcia}}\ and\ \bibinfo {author} {\bibfnamefont {J.}~\bibnamefont
  {Hofmann}},\ }\href@noop {} {\bibinfo {title} {Instanton theory and
  fluctuation corrections to the thermal nucleation rate of a ferromagnetic
  superfluid}} (\bibinfo {year} {2026}),\ \Eprint
  {https://arxiv.org/abs/2512.20734} {arXiv:2512.20734} \BibitemShut {NoStop}%
\bibitem [{\citenamefont {Lagnese}\ \emph {et~al.}(2021)\citenamefont
  {Lagnese}, \citenamefont {Surace}, \citenamefont {Kormos},\ and\
  \citenamefont {Calabrese}}]{Lagnese_21}%
  \BibitemOpen
  \bibfield  {author} {\bibinfo {author} {\bibfnamefont {G.}~\bibnamefont
  {Lagnese}}, \bibinfo {author} {\bibfnamefont {F.~M.}\ \bibnamefont {Surace}},
  \bibinfo {author} {\bibfnamefont {M.}~\bibnamefont {Kormos}},\ and\ \bibinfo
  {author} {\bibfnamefont {P.}~\bibnamefont {Calabrese}},\ }\bibfield  {title}
  {\bibinfo {title} {False vacuum decay in quantum spin chains},\ }\href
  {https://doi.org/10.1103/PhysRevB.104.L201106} {\bibfield  {journal}
  {\bibinfo  {journal} {Phys. Rev. B}\ }\textbf {\bibinfo {volume} {104}},\
  \bibinfo {pages} {L201106} (\bibinfo {year} {2021})}\BibitemShut {NoStop}%
\bibitem [{\citenamefont {Prokof'ev}\ \emph {et~al.}(2001)\citenamefont
  {Prokof'ev}, \citenamefont {Ruebenacker},\ and\ \citenamefont
  {Svistunov}}]{prokofev_01}%
  \BibitemOpen
  \bibfield  {author} {\bibinfo {author} {\bibfnamefont {N.}~\bibnamefont
  {Prokof'ev}}, \bibinfo {author} {\bibfnamefont {O.}~\bibnamefont
  {Ruebenacker}},\ and\ \bibinfo {author} {\bibfnamefont {B.}~\bibnamefont
  {Svistunov}},\ }\bibfield  {title} {\bibinfo {title} {Critical point of a
  weakly interacting two-dimensional {B}ose gas},\ }\href
  {https://doi.org/10.1103/PhysRevLett.87.270402} {\bibfield  {journal}
  {\bibinfo  {journal} {Phys. Rev. Lett.}\ }\textbf {\bibinfo {volume} {87}},\
  \bibinfo {pages} {270402} (\bibinfo {year} {2001})}\BibitemShut {NoStop}%
\bibitem [{\citenamefont {P\^{\i}rvu}\ \emph {et~al.}(2024)\citenamefont
  {P\^{\i}rvu}, \citenamefont {Shkerin},\ and\ \citenamefont
  {Sibiryakov}}]{Pirvu_24}%
  \BibitemOpen
  \bibfield  {author} {\bibinfo {author} {\bibfnamefont {D.}~\bibnamefont
  {P\^{\i}rvu}}, \bibinfo {author} {\bibfnamefont {A.}~\bibnamefont
  {Shkerin}},\ and\ \bibinfo {author} {\bibfnamefont {S.}~\bibnamefont
  {Sibiryakov}},\ }\bibfield  {title} {\bibinfo {title} {Thermal false vacuum
  decay in (1+1) dimensions: Evidence for nonequilibrium dynamics},\ }\href
  {https://doi.org/10.1142/S0217751X24450076} {\bibfield  {journal} {\bibinfo
  {journal} {Int. J. Mod. Phys. A}\ }\textbf {\bibinfo {volume} {39}},\
  \bibinfo {pages} {2445007} (\bibinfo {year} {2024})}\BibitemShut {NoStop}%
\bibitem [{\citenamefont {Billam}\ \emph {et~al.}(2019)\citenamefont {Billam},
  \citenamefont {Gregory}, \citenamefont {Michel},\ and\ \citenamefont
  {Moss}}]{Billam_19}%
  \BibitemOpen
  \bibfield  {author} {\bibinfo {author} {\bibfnamefont {T.~P.}\ \bibnamefont
  {Billam}}, \bibinfo {author} {\bibfnamefont {R.}~\bibnamefont {Gregory}},
  \bibinfo {author} {\bibfnamefont {F.}~\bibnamefont {Michel}},\ and\ \bibinfo
  {author} {\bibfnamefont {I.~G.}\ \bibnamefont {Moss}},\ }\bibfield  {title}
  {\bibinfo {title} {Simulating seeded vacuum decay in a cold atom system},\
  }\href {https://doi.org/10.1103/PhysRevD.100.065016} {\bibfield  {journal}
  {\bibinfo  {journal} {Phys. Rev. D}\ }\textbf {\bibinfo {volume} {100}},\
  \bibinfo {pages} {065016} (\bibinfo {year} {2019})}\BibitemShut {NoStop}%
\bibitem [{Boo()}]{Bootstrapping}%
  \BibitemOpen
  \href@noop {} {}\bibinfo {note} {The decay rate, $\Gamma$ is extracted from
  the survival probability $\langle \cal{Z}$$(t) \rangle$, after averaging over
  individual stochastic realizations. This prevents a direct access to the
  intrinsic statistical distribution of $\Gamma$. Bootstrapping provides a way
  to estimate the uncertainty of the intrinsic distribution from the existing
  dataset ($\mathcal{N}=100$) without generating additional data. From the
  original 100 trajectories, we construct 10,000 bootstrapped samples. For each
  bootstrapped sample, 100 trajectories are drawn with replacement, and
  $\Gamma$ is computed by exponential fitting in the range $\langle
  \cal{Z}$$(t) \rangle \in [0.5, 0.9]$ as shown in the main text. The resulting
  standard deviation of the distribution of $\Gamma$ across the bootstrapped
  samples yield error bars for each temperature. An identical bootstrapping
  protocol is adopted to estimate the errorbars in the decay rates extracted
  from $P(t)$.}\BibitemShut {Stop}%
\bibitem [{\citenamefont {Sz\'asz-Schagrin}\ and\ \citenamefont
  {Tak\'acs}(2022)}]{takacs_22}%
  \BibitemOpen
  \bibfield  {author} {\bibinfo {author} {\bibfnamefont {D.}~\bibnamefont
  {Sz\'asz-Schagrin}}\ and\ \bibinfo {author} {\bibfnamefont {G.}~\bibnamefont
  {Tak\'acs}},\ }\bibfield  {title} {\bibinfo {title} {False vacuum decay in
  the (1+1)-dimensional ${\varphi}^{4}$ theory},\ }\href
  {https://doi.org/10.1103/PhysRevD.106.025008} {\bibfield  {journal} {\bibinfo
   {journal} {Phys. Rev. D}\ }\textbf {\bibinfo {volume} {106}},\ \bibinfo
  {pages} {025008} (\bibinfo {year} {2022})}\BibitemShut {NoStop}%
\bibitem [{\citenamefont {Devoto}\ \emph {et~al.}(2022)\citenamefont {Devoto},
  \citenamefont {Devoto}, \citenamefont {Di~Luzio},\ and\ \citenamefont
  {Ridolfi}}]{Devoto_22}%
  \BibitemOpen
  \bibfield  {author} {\bibinfo {author} {\bibfnamefont {F.}~\bibnamefont
  {Devoto}}, \bibinfo {author} {\bibfnamefont {S.}~\bibnamefont {Devoto}},
  \bibinfo {author} {\bibfnamefont {L.}~\bibnamefont {Di~Luzio}},\ and\
  \bibinfo {author} {\bibfnamefont {G.}~\bibnamefont {Ridolfi}},\ }\bibfield
  {title} {\bibinfo {title} {False vacuum decay: an introductory review},\
  }\href {https://doi.org/10.1088/1361-6471/ac7f24} {\bibfield  {journal}
  {\bibinfo  {journal} {J. Phys. G}\ }\textbf {\bibinfo {volume} {49}},\
  \bibinfo {pages} {103001} (\bibinfo {year} {2022})}\BibitemShut {NoStop}%
\bibitem [{\citenamefont {Berera}\ \emph {et~al.}(2019)\citenamefont {Berera},
  \citenamefont {Mabillard}, \citenamefont {Mintz},\ and\ \citenamefont
  {Ramos}}]{Berera_19}%
  \BibitemOpen
  \bibfield  {author} {\bibinfo {author} {\bibfnamefont {A.}~\bibnamefont
  {Berera}}, \bibinfo {author} {\bibfnamefont {J.}~\bibnamefont {Mabillard}},
  \bibinfo {author} {\bibfnamefont {B.~W.}\ \bibnamefont {Mintz}},\ and\
  \bibinfo {author} {\bibfnamefont {R.~O.}\ \bibnamefont {Ramos}},\ }\bibfield
  {title} {\bibinfo {title} {Formulating the {K}ramers problem in field
  theory},\ }\href {https://doi.org/10.1103/PhysRevD.100.076005} {\bibfield
  {journal} {\bibinfo  {journal} {Phys. Rev. D}\ }\textbf {\bibinfo {volume}
  {100}},\ \bibinfo {pages} {076005} (\bibinfo {year} {2019})}\BibitemShut
  {NoStop}%
\end{thebibliography}%


%

\end{document}